\documentclass[aps,rmp,square]{revtex4}

\usepackage{amsmath,amssymb}
\usepackage{graphicx}
\usepackage{pst-plot}
\usepackage{mathrsfs}
\usepackage{enumerate}
\usepackage{mathtools}
\usepackage{etoolbox}

\def\dbl{\hbox{${1\hskip -2.4pt{\rm l}}$}}

\newtheorem{my-theorem}{Theorem}

\def\ocite#1{[\citenum{#1}]}

\setcitestyle{numbers}

\begin{document}

\title{Refutation of Richard Gill's Argument Against my Disproof of Bell's Theorem}

\author{Joy Christian}

\email{jjc@alum.bu.edu}

\affiliation{Einstein Centre for Local-Realistic Physics, 15 Thackley End, Oxford OX2 6LB, United Kingdom}

\begin{abstract}
I identify a number of errors in Richard Gill's purported refutation of my disproof of Bell's theorem. In particular, I point out that his central argument is based, not only on a rather trivial misreading of my counterexample to Bell's theorem, but also on a simple oversight of a freedom of choice in the orientation of a Clifford algebra. What is innovative and original in my counterexample is thus\break mistaken for an error, at the expense of the professed universality and generality of Bell's theorem.
\end{abstract}

\maketitle

\parskip 5pt

{\baselineskip 11.4235pt

\section{Introduction}

In a recent preprint \ocite{gill}
Richard Gill has suggested that there is an algebraic error in my disproof of Bell's theorem
\ocite{origins}\ocite{disproof}\ocite{Christian-666}\ocite{Restoring}\ocite{What-666}\ocite{illusion-666}\ocite{photon-666}\ocite{reply}\ocite{Further-666}\ocite{experiment-666}\ocite{refutation}. In what follows I show that it is in fact Gill's argument that
is in error, stemming from a rather trivial misreading of my disproof, as well as from a misconstrual of a basic freedom in
geometric algebra. For example, in the abstract of both versions of his preprint he refers to the quantity
${-{\bf a}\cdot{\bf b}-{\bf a}\wedge{\bf b}}$ as a ``bivector.'' Moreover, by his own admission Gill has not read any of my papers
on the subject beyond the minimalist one-page paper \ocite{disproof}. This is unfortunate. For had he read my papers
carefully he would have recognized his errors himself.  

Considering these facts, let me
set the stage by recalling some of the background to my counterexample on which Gill's argument relies \ocite{gill}.
Although based on geometric algebra \ocite{Clifford}, my counterexample is in fact a hidden variable model,
with the hidden variable being the initial orientation of a parallelized 3-sphere \ocite{origins}\ocite{disproof}.
In other words, the initial orientation of the physical space itself is taken as a hidden variable in the model \ocite{What-666}.
With this in mind, consider a right-handed frame of ordered basis bivectors,
${\{{\boldsymbol\beta}_x,\,{\boldsymbol\beta}_y,\,{\boldsymbol\beta}_z\}}$, and the corresponding bivector (or ``even'')
subalgebra
\begin{equation}
{\boldsymbol\beta}_j\,{\boldsymbol\beta}_k \,=\,-\,\delta_{jk}\,-\,\epsilon_{jkl}\,{\boldsymbol\beta}_l\label{lisr}
\end{equation}
of the Clifford algebra ${{Cl}_{3,0}}$ \ocite{Christian-666}.
The latter is a linear vector space ${{\rm I\!R}^8}$ spanned by the orthonormal basis
\begin{equation}
\left\{1,\,\;{\bf e}_x,\,{\bf e}_y,\,{\bf e}_z,\,\;{\bf e}_x\wedge{\bf e}_y,\,
{\bf e}_y\wedge{\bf e}_z,\,{\bf e}_z\wedge{\bf e}_x,\,\;
{\bf e}_x\wedge{\bf e}_y\wedge{\bf e}_z\right\}\!, \label{cl30}
\end{equation}
where ${\delta_{jk}}$ is the Kronecker delta, ${\epsilon_{jkl}}$ is the Levi-Civita symbol, the indices
${j,\,k,\,l=x,\,y,}$ or ${z}$ are cyclic indices, and
\begin{equation}
{\boldsymbol\beta}_j\,=\,{\bf e}_k\wedge{\bf e}_l\,=\,I\cdot{\bf e}_j\,.\label{tog-13}
\end{equation}
Eq.${\,}$(\ref{lisr}) is a conventional expression of bivector subalgebra,
routinely used in the literature employing geometric algebra for computations \ocite{Clifford}.
From equation (\ref{lisr}) it is easy to verify the familiar properties of the basis bivectors, such as
\begin{align}
({\boldsymbol\beta}_x)^2\,=\,({\boldsymbol\beta}_y)^2&\,=\,({\boldsymbol\beta}_z)^2\,=\,-\,1 \label{prop13} \\
\text{and}\;\;\;{\boldsymbol\beta}_x\,{\boldsymbol\beta}_y\,=
&\,-\,{\boldsymbol\beta}_y\,{\boldsymbol\beta}_x\;\;\;\text{etc.} \label{prop14}
\end{align}
Moreover, it is easy to verify that the bivectors satisfying the subalgebra (\ref{lisr})
form a right-handed frame of basis bivectors. To this end, right-multiply both sides of Eq.${\,}$(\ref{lisr}) by
${{\boldsymbol\beta}_l}$, and then use the fact that ${({\boldsymbol\beta}_l)^2=-1}$ to arrive${\;}$at
\begin{equation}
{\boldsymbol\beta}_j\,{\boldsymbol\beta}_k\,{\boldsymbol\beta}_l\,=\,+\,1\,. \label{rightH}
\end{equation}
The fact that this ordered product yields a positive value confirms that
${\{{\boldsymbol\beta}_x,\,{\boldsymbol\beta}_y,\,{\boldsymbol\beta}_z\}}$ indeed forms a right-handed frame of basis bivectors.
This is a universally accepted convention, found in any textbook on geometric algebra \ocite{Clifford}.

Suppose now ${{\bf a}=a_{\! j}\,{\bf e}_j}$ and ${{\bf b}=b_{\! k}\,{\bf e}_k}$ are two unit vectors in ${{\rm I\!R}^3}$,
where the repeated indices are summed over ${x,\,y,}$ and ${z}$.
Then the right-handed basis defined in Eq.${\,}$(\ref{lisr}) leads to
\begin{equation}
\{\,a_j\;{\boldsymbol\beta}_j\,\}\,\{\,b_k\;{\boldsymbol\beta}_k\,\}\,=\,-\,a_j\,b_k\,\delta_{jk}\,
-\,\epsilon_{jkl}\;a_j\,b_k\;{\boldsymbol\beta}_l\,,
\end{equation}
which, together with (\ref{tog-13}), is equivalent to 
\begin{equation}
(\,I\cdot{\bf a})(\,I\cdot{\bf b})\,
=\,-\,{\bf a}\cdot{\bf b}\,-\,I\cdot({\bf a}\times{\bf b}), \label{one-ident}
\end{equation}
where ${I={{\bf e}_x}{{\bf e}_y}{{\bf e}_z}}$ is the standard trivector. Geometrically this identity describes all points of a
parallelized 3-sphere.

Let us now consider a left-handed frame of ordered basis bivectors, which we also denote by
${\{{\boldsymbol\beta}_x,\,{\boldsymbol\beta}_y,\,{\boldsymbol\beta}_z\}}$. It is important to recognize,
however, that there is no {\it a priori} way of knowing that this new basis frame is in fact left-handed.
To ensure that it is indeed left-handed we must first make sure that it is an ordered frame by requiring
that its basis elements satisfy the bivector properties delineated in Eqs.${\,}$(\ref{prop13}) and (\ref{prop14}).
Next, to distinguish this frame from the right-handed frame defined by equation (\ref{rightH}), we must require
that its basis elements satisfy the property
\begin{equation}
{\boldsymbol\beta}_j\,{\boldsymbol\beta}_k\,{\boldsymbol\beta}_l\,=\,-\,1\,. \label{leftH}
\end{equation}
One way to ensure this is to multiply every non-scalar element in (\ref{cl30}) by a minus sign. Then,
instead of (\ref{tog-13}), we have
\begin{equation}
{\boldsymbol\beta}_j\,=\,-\,{\bf e}_k\wedge{\bf e}_l\,=\,(\,-\,I\,)\cdot(\,-\,{\bf e}_j\,)\,=\,I\cdot{\bf e}_j\,,\label{tog-20}
\end{equation}
and the condition (\ref{leftH}) is automatically satisfied.
As is well known, this was the condition imposed by Hamilton on his unit quaternions, which we now know are
nothing but a left-handed set
of basis bivectors \ocite{Clifford}. Indeed, it can be easily checked that the basis bivectors satisfying the properties
(\ref{prop13}), (\ref{prop14}), (\ref{leftH}), and (\ref{tog-20}) compose the subalgebra
\begin{equation}
{\boldsymbol\beta}_j\,{\boldsymbol\beta}_k \,=\,-\,\delta_{jk}\,+\,\epsilon_{jkl}\,{\boldsymbol\beta}_l\,. \label{lisr-left}
\end{equation}
Conversely, it is easy to check that the basis bivectors defined by this subalgebra do indeed form a left-handed frame. To this
end, right-multiply both sides of Eq.${\,}$(\ref{lisr-left}) by ${{\boldsymbol\beta}_l}$, and then use the property
${({\boldsymbol\beta}_l)^2=-1}$ to verify Eq.${\,}$(\ref{leftH}).
As is well known, this subalgebra is generated by the unit quaternions originally proposed by Hamilton \ocite{Clifford}. 
It is routinely used in the textbook treatments of angular momenta, but without mentioning the fact that it defines
nothing but a left-handed set of basis bivectors. It may look more familiar if we temporarily change notation and rewrite
Eq.${\,}$(\ref{lisr-left})${\;}$as
\begin{equation}
{\bf J}_j\,{\bf J}_k \,=\,-\,\delta_{jk}\,+\,\epsilon_{jkl}\,{\bf J}_l\,. \label{lisr-left-J}
\end{equation}
More importantly (and especially since Richard Gill seems to have missed this point), I stress once again that there is no
way to set apart the left-handed frame of basis bivectors from the right-handed frame without appealing to the
intrinsically defined distinguishing conditions (\ref{rightH}) and (\ref{leftH}), or equivalently to the corresponding
subalgebras (\ref{lisr}) and (\ref{lisr-left}). Note also that at no time within my framework the two subalgebras
(\ref{lisr}) and (\ref{lisr-left}) are mixed in any way, either physically or mathematically.
They merely play the role of two distinct and alternative hidden variable possibilities. 

Suppose now ${{\bf a}=a_{\! j}\,{\bf e}_j}$ and ${{\bf b}=b_{\! k}\,{\bf e}_k}$ are two unit vectors in ${{\rm I\!R}^3}$, where
the repeated indices are summed over ${x,\,y,}$ and ${z}$. Then the left-handed basis equation (\ref{lisr-left}) leads to
\begin{equation}
\{\,a_j\;{\boldsymbol\beta}_j\,\}\,\{\,b_k\;{\boldsymbol\beta}_k\,\}\,=\,-\,a_j\,b_k\,\delta_{jk}\,
+\,\epsilon_{jkl}\;a_j\,b_k\;{\boldsymbol\beta}_l
\end{equation}
which, together with (\ref{tog-20}), is equivalent to 
\begin{equation}
(\,I\cdot{\bf a})(\,I\cdot{\bf b})\,
=\,-\,{\bf a}\cdot{\bf b}\,+\,I\cdot({\bf a}\times{\bf b}), \label{two-ident}
\end{equation}
where ${I}$ is the standard trivector. Once again, geometrically this identity describes all points of a
parallelized 3-sphere.

It is important to note, however, that
there is a sign difference in the second term on the RHS of the identities (\ref{one-ident}) and (\ref{two-ident}).
The algebraic meaning of this sign difference is of course clear from the above discussion, and it has been discussed
extensively in most of my papers \ocite{Christian-666}\ocite{Restoring}\ocite{What-666}\ocite{photon-666}\ocite{refutation},
with citations to prior literature \ocite{Eberlein}.
But from the perspective of my model a more important question is: What does this sign difference mean {\it geometrically}${\,}$?
To bring out its geometric meaning, let us rewrite the identities (\ref{one-ident}) and (\ref{two-ident}) as
\begin{equation}
(\,+\,I\cdot{\bf a})(\,+\,I\cdot{\bf b})\,
=\,-\,{\bf a}\cdot{\bf b}\,-\,(\,+\,I\,)\cdot({\bf a}\times{\bf b}) \label{id-1}
\end{equation}
and
\begin{equation}
(\,-\,I\cdot{\bf a})(\,-\,I\cdot{\bf b})\,
=\,-\,{\bf a}\cdot{\bf b}\,-\,(\,-\,I\,)\cdot({\bf a}\times{\bf b}), \label{id-2}
\end{equation}
respectively. The geometrical meaning of the two identities is now transparent if we recall that the bivector
${(\,+\,I\cdot{\bf a})}$ represents a counterclockwise rotation about the ${\bf a}$-axis, whereas the bivector
${(\,-\,I\cdot{\bf a})}$ represents a clockwise rotation about the ${\bf a}$-axis. Accordingly, both identities interrelate
the points of a unit parallelized 3-sphere, but the identity (\ref{id-1}) interrelates points of a positively oriented
3-sphere whereas the identity (\ref{id-2}) interrelates points of a negatively oriented 3-sphere. In other words, the 3-sphere
represented by the identity (\ref{id-1}) is oriented in the counterclockwise sense, whereas the 3-sphere represented by the
identity (\ref{id-2}) is oriented in the clockwise sense. These two alternative orientations of the 3-sphere is then
the random hidden variable ${\lambda=\pm\,1}$ (or the initial state ${\lambda=\pm\,1}$) within my model. 

Given this geometrical picture, it is now easy to appreciate that identity (\ref{id-1}) corresponds to the physical space
characterized by the trivector ${\,+\,I\,}$, whereas identity (\ref{id-2}) corresponds to the physical space characterized
by the trivector ${\,-\,I\,}$ \ocite{Eberlein}. This is further supported by the evident fact that, apart from the choice of
a trivector, the identities (\ref{id-1}) and (\ref{id-2}) represent one and the same subalgebra. Moreover, there is clearly no
{\it a priori} reason for Nature to choose ${\,+\,I\,}$ as a fundamental trivector over ${\,-\,I\,}$. Either choice provides a
perfectly legitimate representation of the physical space, and neither is favored by Nature. Consequently, instead of
characterizing the physical space by fixed basis (\ref{cl30}), we can start out with two alternatively possible
characterizations of the physical space by the {\it hidden} basis
\begin{equation}
\left\{1,\,\;{\bf e}_x,\,{\bf e}_y,\,{\bf e}_z,\,\;{\bf e}_x\wedge{\bf e}_y,\,
{\bf e}_y\wedge{\bf e}_z,\,{\bf e}_z\wedge{\bf e}_x,\,\;\lambda\,
(\,{\bf e}_x\wedge{\bf e}_y\wedge{\bf e}_z\,)\right\}\!, \label{hid-cl30}
\end{equation}
where ${\lambda=\pm\,1}$. Although these considerations and the physical motivations behind them have been
the starting point of my program (see, for example, discussions in Refs.${\,}$\ocite{Christian-666},
\ocite{photon-666}, and \ocite{experiment-666}), Gill seems to have overlooked them completely.

Exploiting the natural freedom of choice in characterizing ${S^3}$ by either ${\,+\,I\,}$ or ${\,-\,I\,}$,
we can now combine the identities (\ref{id-1}) and (\ref{id-2}) into
a single hidden variable equation (at least for the computational purposes):
\begin{equation}
(\,\lambda\,I\cdot{\bf a})(\,\lambda\,I\cdot{\bf b})\,
=\,-\,{\bf a}\cdot{\bf b}\,-\,(\,\lambda\,I\,)\cdot({\bf a}\times{\bf b}), \label{combi-0}
\end{equation}
where ${\lambda=\pm\,1}$ now specifies the orientation of the 3-sphere. It is important to recognize that the difference
between the trivectors ${\,+\,I\,}$ and ${\,-\,I\,}$ in this equation
primarily reflects the difference in the handedness of the bivector
basis ${\{{\boldsymbol\beta}_x,\,{\boldsymbol\beta}_y,\,{\boldsymbol\beta}_z\}}$, and not in the handedness of the vector basis
${\{\,{\bf e}_x,\,{\bf e}_y,\,{\bf e}_z\}}$. This should be evident from the foregone arguments, but let us bring this point
home by considering the following change in the handedness of the vector basis:
\begin{equation}
+\,I\,=\,{{\bf e}_x}{{\bf e}_y}{{\bf e}_z}\;\;\longrightarrow\;\;(+\,{{\bf e}_x})(-\,{{\bf e}_y})(+\,{{\bf e}_z})
\,=\,-\,(\,{{\bf e}_x}{{\bf e}_y}{{\bf e}_z}\,)\,=\,-\,I\,.
\end{equation}
Such a change does not induce a change in the handedness of the bivector basis, since it leaves the product
${{\boldsymbol\beta}_x\,{\boldsymbol\beta}_y\,{\boldsymbol\beta}_z}$ unchanged. This can be easily
verified by recalling that ${{\boldsymbol\beta}_x\equiv I\cdot{\bf e}_x\,}$,
${\;{\boldsymbol\beta}_y\equiv I\cdot{\bf e}_y\,,\,}$ and
${\;{\boldsymbol\beta}_z\equiv I\cdot{\bf e}_z\,,\,}$ and consequently
\begin{equation}
+\,1\,=\,{\boldsymbol\beta}_x\,{\boldsymbol\beta}_y\,{\boldsymbol\beta}_z\;\;\longrightarrow\;\;
(\,-\,I\,)\cdot(+\,{{\bf e}_x})\,(\,-\,I\,)\cdot(-\,{{\bf e}_y})\,(\,-\,I\,)\cdot(+\,{{\bf e}_z})\,=\,
{\boldsymbol\beta}_x\,{\boldsymbol\beta}_y\,{\boldsymbol\beta}_z\,=\,+1\,.
\end{equation}
Conversely, a change in the handedness of bivector basis does not necessarily affect a change in the handedness of vector basis,
but leads instead to
\begin{equation}
+\,1\,=\,{\boldsymbol\beta}_x\,{\boldsymbol\beta}_y\,{\boldsymbol\beta}_z\;\;\longrightarrow\;\;
(\,-\,{\boldsymbol\beta}_x)\,(\,-\,{\boldsymbol\beta}_y)\,(\,-\,{\boldsymbol\beta}_z)\,=\,-\,
(\,{\boldsymbol\beta}_x\,{\boldsymbol\beta}_y\,{\boldsymbol\beta}_z\,)\,=\,-1\,,
\end{equation}
which in turn leads us back to equation (\ref{combi-0}) via equations (\ref{one-ident}) and (\ref{two-ident}). Thus the
sign difference between the trivectors ${\,+\,I\,}$ and ${\,-\,I\,}$ captured in equation (\ref{combi-0}) arises from the
sign difference in the product ${{\boldsymbol\beta}_x\,{\boldsymbol\beta}_y\,{\boldsymbol\beta}_z}$ and not from that in the
product ${{{\bf e}_x}{{\bf e}_y}{{\bf e}_z}}$. It is therefore of very different geometrical significance \ocite{Restoring}.
It corresponds to the difference between two possible orientations of the 3-sphere mentioned above. It is also important
to keep in mind that the combined equation (\ref{combi-0}) is simply a convenient shortcut for representing two completely
independent initial states of the system, one corresponding to the counterclockwise orientation of the 3-sphere and the
other corresponding to the clockwise orientation of the 3-sphere. Moreover, at no time these two alternative
possibilities are mixed during the course of an experiment. They represent two independent physical scenarios,
corresponding to two independent runs\break of the experiment. If we now use the notation ${{\boldsymbol\mu}=\lambda\,I}$,
then the combined identity (\ref{combi-0}) takes the convenient form
\begin{equation}
(\,{\boldsymbol\mu}\cdot{\bf a})(\,{\boldsymbol\mu}\cdot{\bf b})\,
=\,-\,{\bf a}\cdot{\bf b}\,-\,{\boldsymbol\mu}\cdot({\bf a}\times{\bf b}).\label{bi-identitzzzzzzzzzzzz}
\end{equation}

If one remains uncomfortable about using this unconventional identity, then there is always the option of working directly with
the bivector basis themselves, and that is what I do in my one-page paper \ocite{disproof}. Accordingly, let us return to
equations (\ref{lisr}) and (\ref{lisr-left}) and start afresh by writing the basic hidden variable equation of the model as
\begin{equation}
{\boldsymbol\beta}_j(\lambda)\,{\boldsymbol\beta}_k(\lambda)\,=\,-\,\delta_{jk}\,-\,\epsilon_{jkl}\,
{\boldsymbol\beta}_l(\lambda)\,,\label{gill-1}
\end{equation}
with ${{\boldsymbol\beta}_j(\lambda)=\lambda\,{\boldsymbol\beta}_j}$ and ${\lambda=\pm\,1}$ as a fair
coin representing the two possible orientations of the 3-sphere. Note that upon substituting
${{\boldsymbol\beta}_j(\lambda)=\lambda\,{\boldsymbol\beta}_j}$ and using ${\lambda^2=+1}$
the above equation can also be written as
\begin{equation}
{\boldsymbol\beta}_j\,{\boldsymbol\beta}_k\,=\,-\,\delta_{jk}\,-\,\lambda\;\epsilon_{jkl}\,
{\boldsymbol\beta}_l\,,\label{gill-2}
\end{equation}
which in turn, for ${\lambda=+1}$, specializes to 
\begin{equation}
{\boldsymbol\beta}_j\,{\boldsymbol\beta}_k\,=\,-\,\delta_{jk}\,-\,\epsilon_{jkl}\,
{\boldsymbol\beta}_l\,.\label{gill-0}
\end{equation}
As elementary as they are, the last three equations are a major source of confusion in Gill's preprint.

To see the equivalence of these equations with the hidden variable identity (\ref{bi-identitzzzzzzzzzzzz}),
let ${{\bf a}=a_{\! j}\,{\bf e}_j}$ and ${{\bf b}=b_{\! k}\,{\bf e}_k}$ be two unit vectors in ${{\rm I\!R}^3}$.
Using either equation (\ref{gill-2}) or (\ref{gill-1}) and
${{\boldsymbol\beta}_j(\lambda)=\lambda\,{\boldsymbol\beta}_j}$ we then have
\begin{equation}
\{\,a_j\;{\boldsymbol\beta}_j(\lambda)\,\}\,\{\,b_k\;{\boldsymbol\beta}_k(\lambda)\,\}\,=\,
\{\,\lambda\;a_j\;{\boldsymbol\beta}_j\,\}\,\{\,\lambda\;b_k\;{\boldsymbol\beta}_k\,\}\,=\,-\,a_j\,b_k\,\delta_{jk}\,
-\,\lambda\;\epsilon_{jkl}\;a_j\,b_k\;{\boldsymbol\beta}_l\,,
\end{equation}
which is equivalent to the identity (\ref{bi-identitzzzzzzzzzzzz}) with 
${(\,{\boldsymbol\mu}\cdot{\bf a})\equiv\{\,\lambda\;a_j\;{\boldsymbol\beta}_j\,\}}$,
${\;{\boldsymbol\mu}\cdot({\bf a}\times{\bf b})\equiv\{\,\lambda\;\epsilon_{jkl}\;a_j\,b_k\;{\boldsymbol\beta}_l\,\}}$, etc.

Next, let us define the measurement results observed by Alice and Bob within ${S^3}$ as
\begin{align}
S^3\ni{\mathscr A}({\bf a},\,{\lambda})\,=\,\{-\,a_j\;{\boldsymbol\beta}_j\,\}\,\{\,a_k\;{\boldsymbol\beta}_k(\lambda)\,\}
\,=\,(\,-I\cdot{\bf a})(\,{\boldsymbol\mu}\cdot{\bf a})\,&=\,
\begin{cases}
+\,1\;\;\;\;\;{\rm if} &\lambda\,=\,+\,1 \\
-\,1\;\;\;\;\;{\rm if} &\lambda\,=\,-\,1
\end{cases} \label{27gill}
\end{align}
and
\begin{align}
\;\;S^3\ni\,{\mathscr B}({\bf b},\,{\lambda})\,=\,\{\,b_j\;{\boldsymbol\beta}_j(\lambda)\,\}\,\{+\,b_k\;{\boldsymbol\beta}_k\,\}
\,=\,(\,{\boldsymbol\mu}\cdot{\bf b})(\,+I\cdot{\bf b})\,&=\,
\begin{cases}
-\,1\;\;\;\;\;{\rm if} &\lambda\,=\,+\,1 \\
+\,1\;\;\;\;\;{\rm if} &\lambda\,=\,-\,1\,.
\end{cases} \label{28gill}
\end{align}
It is important to note that these measurement results are generated as products of two numbers,
the fixed bivectors ${(\,I\cdot{\bf n})}$ times the random bivectors ${(\,{\boldsymbol\mu}\cdot{\bf n})}$. In other words, as
discussed in greater detail in Refs.${\,}$\ocite{origins} and \ocite{Restoring},
they are generated with {\it different} bivectorial scales of dispersion for each measurement
directions ${\bf a}$ and ${\bf b}$. Consequently, as discovered by Galton and Pearson \ocite{scores-2} over a century ago,
the correct correlation between the raw numbers ${\mathscr A}$
and ${\mathscr B}$ (as observed and manipulated by experimentalists) can only be inferred -- {\it theoretically} --
by calculating the covariance of the corresponding standardized variables
${{\boldsymbol\mu}\cdot{\bf a}}$ and ${{\boldsymbol\mu}\cdot{\bf b}\,}$:
\begin{align}
{\cal E}({\bf a},\,{\bf b})\,=\lim_{\,n\,\gg\,1}\left[\frac{1}{n}\sum_{i\,=\,1}^{n}\,
{\mathscr A}({\bf a},\,{\boldsymbol\mu}^i)\;{\mathscr B}({\bf b},\,{\boldsymbol\mu}^i)\right]
\,=\lim_{\,n\,\gg\,1}\left[\frac{1}{n}\sum_{i\,=\,1}^{n}\,
(\,{\boldsymbol\mu}^i\cdot{\bf a})(\,{\boldsymbol\mu}^i\cdot{\bf b})\right]
\,=\,-\,{\bf a}\cdot{\bf b}\,.
\end{align}
Here the last equality immediately follows from the identity (\ref{bi-identitzzzzzzzzzzzz}), and the standardized variables
are calculated as
\begin{align}
A({\bf a},\,{\boldsymbol\mu})&=\frac{\,{\mathscr A}({\bf a},\,{\boldsymbol\mu})\,-\,
{\overline{{\mathscr A}({\bf a},\,{\boldsymbol\mu})}}}{\sigma({\mathscr A})} \notag \\
\,&=\,\frac{\,{\mathscr A}({\bf a},\,{\boldsymbol\mu})\,-\,0\,}{(-\,I\cdot{{\bf a}}\,)}
\,=\,(\,+\,{\boldsymbol\mu}\cdot{{\bf a}}\,) \label{var-a}
\end{align}
and
\begin{align}
B({\bf b},\,{\boldsymbol\mu})&=\frac{\,{\mathscr B}({\bf b},\,{\boldsymbol\mu})\,-\,
{\overline{{\mathscr B}({\bf b},\,{\boldsymbol\mu})}}}{\sigma({\mathscr B})} \notag \\
\,&=\,\frac{\,{\mathscr B}({\bf b},\,{\boldsymbol\mu})\,-\,0\,}{(+\,I\cdot{{\bf b}}\,)}
\,=\,(\,+\,{\boldsymbol\mu}\cdot{{\bf b}}\,),\label{var-b}
\end{align}
with ${\sigma({\mathscr A})=(-\,I\cdot{{\bf a}}\,)}$ and ${\sigma({\mathscr B})=(+\,I\cdot{{\bf b}}\,)}$,
respectively, being the standard deviations in the results ${\mathscr A}$ and ${\mathscr B}$.
The above result may be seen more transparently by recalling
that ${(\,{\boldsymbol\mu}\cdot{\bf a})\equiv\{\,\lambda\;a_j\;{\boldsymbol\beta}_j\,\}}$
and ${(\,{\boldsymbol\mu}\cdot{\bf b})\equiv\{\,\lambda\;b_k\;{\boldsymbol\beta}_k\,\}}$, so that
\begin{equation}
\lim_{n\,\gg\,1}\left[\frac{1}{n}\sum_{i\,=\,1}^{n}\,                                                                     
\left\{\,\lambda^i\;a_j\;{\boldsymbol\beta}_j\,\right\}\,\left\{\,\lambda^i\;b_k\;{\boldsymbol\beta}_k\,\right\}\right]
=\,-\,a_j\,b_j\,-\lim_{n\,\gg\,1}\left[\frac{1}{n}\sum_{i\,=\,1}^{n}\,                                                   
\left\{\,\lambda^i\,\epsilon_{jkl}\;a_j\,b_k\;{\boldsymbol\beta}_l\,\right\}\right]
=\,-\,a_j\,b_j\,+\,0\,=\,-\,{\bf a}\cdot{\bf b}\,, \label{corre-one}
\end{equation}
\begin{equation}
\text{and}\;\;\;\;\;\;\;\;\;\;\;\;\;\;\;\lim_{n\,\gg\,1}\left[\frac{1}{n}\sum_{i\,=\,1}^{n}\,
\left\{\,\lambda^i\,a_j\;{\boldsymbol\beta}_j\,\right\}\right]\,=\,0\,=\,
\lim_{n\,\gg\,1}\left[\frac{1}{n}\sum_{i\,=\,1}^{n}\,\left\{\,\lambda^i\,b_k\;{\boldsymbol\beta}_k\,\right\}\right].
\;\;\;\;\;\;\;\;\;\;\;\;\;\;\;\;\;\;\text{} \label{corre-two}
\end{equation}
It is important to remember that what is being summed over here are points of a parallelized 3-sphere representing
the outcomes of completely independent experimental runs in an EPR-Bohm experiment.
In statistical terms what these results are then showing is that correlation between the raw numbers
${{\mathscr A}({\bf a},\,{\boldsymbol\mu})\,=\,(-\,I\cdot{{\bf a}}\,)
\,(\,+\,{\boldsymbol\mu}\cdot{{\bf a}}\,)\,=\,\pm\,1\in S^3}$ and
${{\mathscr B}({\bf b},\,{\boldsymbol\mu})\,=\,(+\,I\cdot{{\bf b}}\,)
\,(\,+\,{\boldsymbol\mu}\cdot{{\bf b}}\,)\,=\,\pm\,1\in S^3}$
is ${-\,{\bf a}\cdot{\bf b}}$. According to Bell's theorem this is mathematically impossible. 
Further physical, mathematical, and statistical details of this ``impossible'' result can be found in
Refs.${\,}$\ocite{disproof} and \ocite{Restoring}.

\section{A Fallacy of Misplaced Concreteness\protect\footnotemark}\footnotetext{A fallacy of neglecting the degree of
abstraction involved in a thought that leads to an unwarranted conclusion about a concrete entity.}

With this background, we are now in a position to appreciate where the confusion in Gill's argument stems from. To begin
with, he has failed to understand what the hidden variable ${\lambda}$ is in my model. He does not seem to realize that
rejecting the ${\lambda}$ described above as a hidden variable (for that is what his argument boils down to) is equivalent
to rejecting the professed universality and generality of Bell's theorem. This is, however, not so easy to see if one insists on
neglecting my substantive papers on the subject and concentrates solely on the minimalist one-page paper.

Secondly,
Gill has missed the reciprocal relation between the two sets of basis defined in equations (\ref{gill-1}) and  (\ref{gill-0}):
\begin{equation}
{\boldsymbol\beta}_j(\lambda)\,=\,\lambda\,{\boldsymbol\beta}_j\iff
{\boldsymbol\beta}_j\,=\,\lambda\,{\boldsymbol\beta}_j(\lambda).\label{gill-3}
\end{equation}
This relation holds simply because ${\lambda=\pm\,1}$ implies ${\lambda^2=+1}$. 
Just as the relation ${{\boldsymbol\beta}_j(\lambda)\,=\,\lambda\,{\boldsymbol\beta}_j}$ 
encapsulates the randomness of the basis
${\{{\boldsymbol\beta}_j(\lambda)\}}$ with respect to the basis ${\{{\boldsymbol\beta}_j\}}$,
the reciprocal relation ${{\boldsymbol\beta}_j\,=\,\lambda\,{\boldsymbol\beta}_j(\lambda)}$
encapsulates the randomness of the basis
${\{{\boldsymbol\beta}_j\}}$ with respect to the basis ${\{{\boldsymbol\beta}_j(\lambda)\}}$.
As we shall soon see, oversight of this very simple reciprocal relationship between the bases
${\{{\boldsymbol\beta}_j\}}$ and ${\{{\boldsymbol\beta}_j(\lambda)\}}$
invalidates the central contention of Gill rather trivially.

The third source of confusion in Gill's preprint is the equation (\ref{gill-1}) defined above, namely
\begin{equation}
{\boldsymbol\beta}_j(\lambda)\,{\boldsymbol\beta}_k(\lambda)\,=\,-\,\delta_{jk}\,-\,\epsilon_{jkl}\,
{\boldsymbol\beta}_l(\lambda)\,.\label{gill-11}
\end{equation}
This equation, which purportedly appears as equation (2) in his preprint, is {\it incorrectly} presented therein as
\begin{equation}
{\boldsymbol\beta}_j(\lambda)\,{\boldsymbol\beta}_k(\lambda)\,=\,-\,\delta_{jk}\,-\,\lambda\;\epsilon_{jkl}\,
{\boldsymbol\beta}_l(\lambda)\,.\label{bad-gill}
\end{equation}
He attributes this equation to me, but it has nothing to do with my model. Gill's mistake here is not as innocuous as it may
seem at first sight. What he has failed to recognize is that in the former equation ${\lambda}$ represents a hidden variable
({\it i.e.}, the initial orientation of a parallelized 3-sphere), whereas in the latter equation it represents a mere convention.
This reveals that Gill has {\it fundamentally} misunderstood my model \ocite{Christian-666}.
Thus, mistaken reading of (\ref{gill-11}), neglect of the significance of (\ref{gill-3}), neglect of all my substantive
papers but the one-page summary paper, and unfamiliarity with elementary Clifford algebra seems to have led Gill to erroneously
conclude that there is an error in my paper.

There is yet another reason why Gill has been led to this fallacious conclusion. Deviating from the explicit physical model
described above, he insists on unnaturally identifying the actually observed numbers ${\mathscr A}$ and ${\mathscr B}$
with the hidden variable ${\lambda}$ by {\it illegally} treating ${{\mathscr A}({\bf a},\,{\lambda})}$ and
${{\mathscr B}({\bf b},\,{\lambda})}$ as purely algebraic variables rather than statistical variables.
But such an identification is anathema, not only from the physical point of view, but also from the statistical point of view.
${{\mathscr A}({\bf a},\,{\lambda})}$ and ${{\mathscr B}({\bf b},\,{\lambda})}$ are two {\it different} functions of the
random variable ${\lambda}$. What is more, they necessarily describe two {\it statistically independent events} occurring
within a parallelized 3-sphere. Therefore the joint probability of their occurrence
is given by ${P({\mathscr A}\;\text{and}\;{\mathscr B})=P({\mathscr A})\times P({\mathscr B})\leq\frac{1}{2}}$.
And their product, ${{\mathscr A}{\mathscr B}({\bf a},\,{\bf b},\,{\lambda})}$, which itself is necessarily
a {\it different} random variable, is guaranteed to be equal to ${-1}$ only for the case ${{\bf a}={\bf b}}$. For all other
${\bf a}$ and ${\bf b}$, ${{\mathscr A}{\mathscr B}}$ will inevitably alternate between the values ${-1\;\text{and}\;+1}$,
since the numbers ${\mathscr A}$ and ${\mathscr B}$ are being generated with {\it different} bivectorial scales of dispersion.
This is evidently confirmed by the correlation ${-{\bf a}\cdot{\bf b}}$ derived in Eq.${\,}$(\ref{corre-one}).

Notwithstanding, let us play along Gill's illegal game to see where it leads. Let us unnaturally and unphysically set
\begin{equation}
{\mathscr A}({\bf a},\,{\lambda})\,=\,+\,\lambda \;\;\;\;\text{and}\;\;\;\;
{\mathscr B}({\bf b},\,{\lambda})\,=\,-\,\lambda
\end{equation}
so that ${{\mathscr A}{\mathscr B}\,=\,-\,1}$ for all ${\bf a}$ and ${\bf b}$. Now, to begin with, this immediately leads
to ${\sigma({\mathscr A})=\sigma({\mathscr B})=+1}$. This is yet another indication that the above identification is not
only illegal, but has also nothing to do with my model. What is more, since the unnatural identification inevitably leads
to the conclusion that ${{\mathscr A}\,{\mathscr B}\,=\,-\,1}$ for all ${\bf a}$ and ${\bf b}$, it is at variance, not only
with the basic rules of statistical inference, but also with the basic topological properties of the
3-sphere \ocite{origins}\ocite{Restoring}\ocite{What-666}. More specifically, what Gill has failed to recognize is that,
as noted above, ${{\mathscr A}({\bf a},\,{\lambda})}$ and ${{\mathscr B}({\bf b},\,{\lambda})}$
are generated within my model with {\it different} bivectorial scales of dispersion, and
hence the correct correlation between them can be inferred only by calculating the covariation of the
corresponding standardized variables ${{\boldsymbol\mu}\cdot{\bf a}}$ and ${{\boldsymbol\mu}\cdot{\bf b}\,}$:
\begin{align}
{\cal E}({\bf a},\,{\bf b})\,=\lim_{\,n\,\gg\,1}\left[\frac{1}{n}\sum_{i\,=\,1}^{n}\,
{\mathscr A}({\bf a},\,{\boldsymbol\mu}^i)\;{\mathscr B}({\bf b},\,{\boldsymbol\mu}^i)\right]
\,=\lim_{\,n\,\gg\,1}\left[\frac{1}{n}\sum_{i\,=\,1}^{n}\,
(\,{\boldsymbol\mu}^i\cdot{\bf a})(\,{\boldsymbol\mu}^i\cdot{\bf b})\right]
\,=\,-\,{\bf a}\cdot{\bf b}\,.
\end{align}
I have explained the relationship between raw scores and standard scores in greater detail in Ref.${\,}$\ocite{Restoring}, with
explicit calculations for the optical EPR correlations observed in both Orsay and Innsbruck experiments \ocite{Nature-666}.

But I am digressing. Let us continue to play along Gill's unnatural game and recalculate the correlation between
${\mathscr A}$ and ${\mathscr B}$. For completeness, let us first rewrite how this correlation has been calculated in my one-page
paper \ocite{disproof}:
\begin{align}                                                                                                                        
{\cal E}({\bf a},\,{\bf b})\,&=\;\frac{\lim_{\,n\,\gg\,1}\left\{\frac{1}{n}\sum_{i\,=\,1}^{n}\,{\mathscr A}({\bf a},\,{\lambda}^i)\,{\mathscr B}(\
{\bf b},\,{\lambda}^i)\right\}}{\{-\,a_j\;{\boldsymbol\beta}_j\,\}\,\{\,b_k\;{\boldsymbol\beta}_k\,\}}                               
\;=\,\lim_{\,n\,\gg\,1}\left[\frac{1}{n}\sum_{i\,=\,1}^{n}\,\frac{{\mathscr A}({\bf a},\,{\lambda}^i)\,{\mathscr B}({\bf b},\,{\lambda}^i)}{\{-\,\
a_j\;{\boldsymbol\beta}_j\,\}\,\{\,b_k\;{\boldsymbol\beta}_k\,\}}\right] \\                                                
&=\lim_{\,n\,\gg\,1}\left[\frac{1}{n}\sum_{i\,=\,1}^{n}\,\{\,a_j\;{\boldsymbol\beta}_j\,\}                                  
\left\{\,{\mathscr A}({\bf a},\,{\lambda}^i)\,{\mathscr B}({\bf b},\,{\lambda}^i)\right\}\{-\,b_k\;{\boldsymbol\beta}_k\,\}\right] \,=                     
\lim_{\,n\,\gg\,1}\left[\frac{1}{n}\sum_{i\,=\,1}^{n}\,                                                                     
\left\{\,a_j\;{\boldsymbol\beta}_j(\lambda^i)\,\right\}\,\left\{\,b_k\;{\boldsymbol\beta}_k(\lambda^i)\,\right\}\right]  \\          
&=\,-\,a_j\,b_j\,-\lim_{\,n\,\gg\,1}\left[\frac{1}{n}\sum_{i\,=\,1}^{n}\,                                                   
\left\{\,\lambda^i\,\epsilon_{jkl}\;a_j\,b_k\;{\boldsymbol\beta}_l\,\right\}\right]                                                  
=\,-\,a_j\,b_j\,+\,0\,=\,-\,{\bf a}\cdot{\bf b}\,,                                                                                   
\end{align}
where I have used Eq.${\,}$(\ref{gill-1}) and ${{\boldsymbol\beta}_j(\lambda)\,=\,\lambda\,{\boldsymbol\beta}_j}$
from Eq.${\,}$(\ref{gill-3}) in the last two lines of the derivation.

Now, complying with Gill's specious demand, if we set ${{\mathscr A}{\mathscr B}\,=\,-\,1}$
for all ${\bf a}$ and ${\bf b}$, then instead of the above derivation we have the following:
\begin{align}
{\cal E}({\bf a},\,{\bf b})\,&=\;\frac{\lim_{\,n\,\gg\,1}\left\{\frac{1}{n}\sum_{i\,=\,1}^{n}\,{\mathscr A}({\bf a},\,{\lambda}^i)\,{\mathscr B}(\
{\bf b},\,{\lambda}^i)\right\}}{\{-\,a_j\;{\boldsymbol\beta}_j\,\}\,\{\,b_k\;{\boldsymbol\beta}_k\,\}}                               
\;=\,\lim_{\,n\,\gg\,1}\left[\frac{1}{n}\sum_{i\,=\,1}^{n}\,\frac{-1}{\{-\,\
a_j\;{\boldsymbol\beta}_j\,\}\,\{\,b_k\;{\boldsymbol\beta}_k\,\}}\right] \label{5} \\                                                
&=\lim_{\,n\,\gg\,1}\left[\frac{1}{n}\sum_{i\,=\,1}^{n}\,\{\,a_j\;{\boldsymbol\beta}_j\,\}                                  
\left\{-1\right\}\{-\,b_k\;{\boldsymbol\beta}_k\,\}\right] \,=                     
\lim_{\,n\,\gg\,1}\left[\frac{1}{n}\sum_{i\,=\,1}^{n}\,                                                                     
\left\{\,a_j\;{\boldsymbol\beta}_j\,\right\}\,\left\{\,b_k\;{\boldsymbol\beta}_k\,\right\}\right]  \\          
&=\,-\,a_j\,b_j\,-\lim_{\,n\,\gg\,1}\left[\frac{1}{n}\sum_{i\,=\,1}^{n}\,                                                   
\left\{\,\lambda^i\,\epsilon_{jkl}\;a_j\,b_k\;{\boldsymbol\beta}_l(\lambda)\,\right\}\right]                                                  
=\,-\,a_j\,b_j\,+\,0\,=\,-\,{\bf a}\cdot{\bf b}\,, \label{corgill}                                                                                  
\end{align}
where I have used Eq.${\,}$(\ref{gill-0}) and ${{\boldsymbol\beta}_j=\lambda\,{\boldsymbol\beta}_j(\lambda)}$
from Eq.${\,}$(\ref{gill-3}) in the last two lines of derivation, which is equivalent to using
Eq.${\,}$(\ref{gill-2}). Evidently, the correlation between ${\mathscr A}$ and ${\mathscr B}$ does not change,
even from this unnatural perspective.

One may wonder, however, about my use of the relation
${{\boldsymbol\beta}_j\,=\,\lambda\,{\boldsymbol\beta}_j(\lambda)}$
in this derivation in addition to using Eq.${\,}$(\ref{gill-0}).
After all, are not the basis ${\{{\boldsymbol\beta}_x,\,{\boldsymbol\beta}_y,\,{\boldsymbol\beta}_z\}}$ supposed to
be the ``fixed'' bivector basis and the basis
${\{{\boldsymbol\beta}_x(\lambda),\,{\boldsymbol\beta}_y(\lambda),\,{\boldsymbol\beta}_z(\lambda)\}}$
dependent on the hidden variable ${\lambda}$? We must not forget what we did, however, in the first line of the
derivation, in order to arrive at the last two lines. Complying with Gill's specious demand,
we artificially treated the statistical variables
${{\mathscr A}({\bf a},\,{\lambda})}$ and ${{\mathscr B}({\bf b},\,{\lambda})}$ as if they were
purely algebraic variables, and forced the value of their product ${{\mathscr A}{\mathscr B}({\bf a},\,{\bf b},\,{\lambda})}$
to be equal to ${-1}$ for all ${\bf a}$ and ${\bf b}$. In other words,
we surreptitiously assumed that the numbers ${\mathscr A}$ and ${\mathscr B}$ are completely dispersion-free. But
how can that be? Given their definitions (\ref{27gill}) and (\ref{28gill}), it is clear that ${\mathscr A}$ and ${\mathscr B}$
are {\it not} generated as dispersion-free numbers, at least from the perspectives of the detectors
${(\,-I\cdot{\bf a})}$ and ${(\,+I\cdot{\bf b})}$:
\begin{align}
{\mathscr A}\,&=\,(\,-I\cdot{\bf a})(\,{\boldsymbol\mu}\cdot{\bf a}) \\
\text{and}\;\;\;{\mathscr B}\,&=\,(\,{\boldsymbol\mu}\cdot{\bf b})(\,+I\cdot{\bf b}).
\end{align}
The scalar ${\mathscr A}$ is generated with a bivectorial scale of dispersion ${(\,-I\cdot{\bf a})}$ due to randomness
within ${(\,{\boldsymbol\mu}\cdot{\bf a})}$, and the scalar ${\mathscr B}$ is generated with a bivectorial scale of
dispersion ${(\,+I\cdot{\bf b})}$ due to randomness within ${(\,{\boldsymbol\mu}\cdot{\bf b})}$.
Thus, clearly, ${\mathscr A}$ and ${\mathscr B}$ can be treated as dispersion-free {\it only} from the perspectives
of the spins ${(\,{\boldsymbol\mu}\cdot{\bf a})}$ and ${(\,{\boldsymbol\mu}\cdot{\bf b})}$ themselves
rather than from those of the detectors ${(\,-I\cdot{\bf a})}$ and ${(\,+I\cdot{\bf b})}$. In other words, fixing the
value of the product ${{\mathscr A}{\mathscr B}}$ amounts to
viewing the correlation between ${\mathscr A}$ and ${\mathscr B}$ from the perspectives of the spins rather than those of
the detectors \ocite{origins}. But from the perspectives of the spins the detectors are {\it not}
fixed but alternate their handedness between left
and right, precisely as dictated by the relation ${{\boldsymbol\beta}_j\,=\,\lambda\,{\boldsymbol\beta}_j(\lambda)}$
we have used in deriving the correlation (\ref{corgill}).
And as we discussed above, this relation encapsulates the randomness within ${(\,-I\cdot{\bf a})}$
{\it relative} to the spin ${(\,{\boldsymbol\mu}\cdot{\bf a})}$. Moreover, within my model the scalars
${\mathscr A}$, ${\mathscr B}$, and ${{\mathscr A}{\mathscr B}}$
and the bivectors ${(\,-I\cdot{\bf a})}$, ${(\,+I\cdot{\bf b})}$,
${(\,{\boldsymbol\mu}\cdot{\bf a})}$, and ${(\,{\boldsymbol\mu}\cdot{\bf b})}$ are all supposed to be
different points of a parallelized 3-sphere. Thus the above derivation once again reinforces the
central view of my program that EPR
correlations are nothing but correlations among the points of a parallelized 3-sphere, regardless of its algebraic
representation. Consequently, it is not at all surprising that algebraic consistency continues to hold between
the natural perspective advocated in Refs.${\,}$\ocite{origins} and \ocite{Restoring}
and the unnatural perspective insisted upon by Gill.

\section{Conclusion}

The argument of Gill against my disproof of Bell's theorem is based on what Whitehead would have called a
fallacy of\break misplaced concreteness.
Instead of trying to understand my model from a natural, physical
perspective, Gill insists on understanding it from an abstract, unphysical perspective, and claims that that
exposes an algebraic error in my paper. However, it turns out that even this misplaced strategy fails, because
my model passes his peculiar algebraic test with flying colors. Put differently, Gill's argument is misguided on
more than one counts. To begin with, it is based on a trivial misreading of my local-realistic model, as well as on an
oversight of a freedom of choice in the orientation of a parallelized 3-sphere.
In addition to this, there are a number of elementary
mathematical errors in his argument which by themselves are sufficient to undermine his conclusions. Upon
his request I have tried to ignore these errors and concentrate on his central argument only. It is however
worth noting that, trivial as they may be, the errors start building up right from the abstract of his preprint.
For instance, in his abstract he states that ``Correctly computed, [my] standardized correlation are the bivectors
${-{\bf a}\cdot{\bf b}-{\bf a}\wedge{\bf b}\dots}$'' This statement is nonsensical even in the corrected\break second
version of his preprint. More importantly, I have shown that Gill's argument stems from an erroneous reading
of my central equation (\ref{gill-1}), not recognizing the significance of the reciprocity relation (\ref{gill-3})
implicit in my papers, not reading my substantive papers but only the one-page summary paper,
and unfamiliarity with basic Clifford algebra. This leads him to erroneously conclude that there is an error
in my paper. I hope I have succeeded in demonstrating that this conclusion is false. More specifically, the
EPR correlation predicted by my local-realistic model are precisely
\begin{align}
{\cal E}({\bf a},\,{\bf b})\,=\lim_{\,n\,\gg\,1}\left[\frac{1}{n}\sum_{i\,=\,1}^{n}\,
{\mathscr A}({\bf a},\,{\boldsymbol\mu}^i)\;{\mathscr B}({\bf b},\,{\boldsymbol\mu}^i)\right]
\,=\,-\,{\bf a}\cdot{\bf b}\,,
\end{align}
where ${S^3\ni{\mathscr A}({\bf a},\,{\boldsymbol\mu})=\pm\,1}$ and
${S^3\ni{\mathscr B}({\bf b},\,{\boldsymbol\mu})=\pm\,1}$ are the unadorned raw scores observed by Alice and Bob.
Given this clear and straightforward result one may wonder why Gill ends up getting a different result. The
answer is quite simple. He is working with a counterfeit of my model, with little or no incentive to understand
the real model. It is therefore not all that surprising that he ends up getting the result
${{\cal E}({\bf a},\,{\bf b})=\text{nonsense}}$, instead of ${{\cal E}({\bf a},\,{\bf b})= -\,{\bf a}\cdot{\bf b}}$.

Finally, it is worth noting that, contrary to Gill's failed strategy, my model \ocite{disproof} is based on a substantive physical
hypothesis. I hypothesize that the space we live in respects the symmetries and topologies of a parallelized 3-sphere,
which is one of the infinitely many fibers of a parallelized 7-sphere. The EPR correlations are thus correlations
among the points of a parallelized 3-sphere, whereas quantum correlations in general are
correlations among the point of a parallelized 7-sphere. My program thus goes far deeper and well
beyond the narrow confines of Bell's theorem \ocite{origins}.

\vspace{-0.15cm}

\acknowledgments

I wish to thank Richard Gill for nearly a month-long correspondence about my one-page paper, building up to his critique.
I regret not having persuaded him so far, but hope to do better with this formal response to his critique.
I also wish to thank the Foundational Questions Institute (FQXi) for supporting this work through a Mini-Grant.

}

\renewcommand{\bibnumfmt}[1]{[{\bf A}#1]}
\renewcommand{\citenumfont}[1]{{\bf A}#1}

\appendix
\section{\large Refutation of Richard Gill's New Argument Against my Disproof of Bell's Theorem}

Richard Gill has suggested online that the derivation of the identities (\ref{one-ident}) and (\ref{two-ident}) above can be questioned because of my use of the same notation ${{\boldsymbol\beta}_i}$ for both the right- and left-handed bivectors. To dispel any such doubt, let me rederive\break those identities here more carefully using different notations for the right- and the left-handed bivectors. To this end, consider a right-handed frame of ordered basis bivectors, ${\{{\boldsymbol\alpha}_x,\,{\boldsymbol\alpha}_y,\,{\boldsymbol\alpha}_z\}}$, and the corresponding bivector sub-algebra
\begin{equation}
{\boldsymbol\alpha}_i\,{\boldsymbol\alpha}_j \,=\,-\,\delta_{ij}\,-\,\epsilon_{ijk}\,{\boldsymbol\alpha}_k\label{Alisr}
\end{equation}
of the Clifford algebra ${{Cl}_{3,0}}$. The latter is a vector space, ${{\rm I\!R}^8}$, spanned by the ordered set of graded orthonormal basis
\begin{equation}
\left\{1,\,\;{\bf e}_x,\,{\bf e}_y,\,{\bf e}_z,\,\;
{\bf e}_y\wedge{\bf e}_z,\,{\bf e}_z\wedge{\bf e}_x,\,{\bf e}_x\wedge{\bf e}_y,\,\;
{\bf e}_x\wedge{\bf e}_y\wedge{\bf e}_z\right\}\!, \label{Acl30}
\end{equation}
where ${\delta_{ij}}$ is the Kronecker delta, ${\epsilon_{ijk}}$ is the Levi-Civita symbol, the indices ${i,\,j,\,k=x,\,y,}$ or ${z}$ are cyclic indices, and
\begin{equation}
{\boldsymbol\alpha}_i\,=\,{\bf e}_j\wedge{\bf e}_k\,=\,I\cdot{\bf e}_i\,,\label{Atog-13}
\end{equation}
with ${I:={\bf e}_x\wedge{\bf e}_y\wedge{\bf e}_z}$ being a volume form of physical space. Eq.${\,}$(\ref{Alisr}) is a standard definition of bivector subalgebra, routinely used in geometric algebra \ocite{GA}. From it, it is easy to verify the basic properties of the basis bivectors,${\;}$such${\;}$as
\begin{align}
({\boldsymbol\alpha}_x)^2\,=\,({\boldsymbol\alpha}_y)^2&\,=\,({\boldsymbol\alpha}_z)^2\,=\,-\,1 \label{Aprop13} \\
\text{and}\;\;\;{\boldsymbol\alpha}_x\,{\boldsymbol\alpha}_y\,=
&\,-\,{\boldsymbol\alpha}_y\,{\boldsymbol\alpha}_x\;\;\;\text{etc.} \label{Aprop14}
\end{align}
Moreover, it is easy to verify that the bivectors satisfying the subalgebra (\ref{Alisr})
form a right-handed frame of basis bivectors. To check this, right-multiply both sides of Eq.${\,}$(\ref{Alisr}) by ${{\boldsymbol\alpha}_k}$, and then use the fact that ${({\boldsymbol\alpha}_k)^2=-1\;}$to${\;}$arrive${\;}$at
\begin{equation}
{\boldsymbol\alpha}_i\,{\boldsymbol\alpha}_j\,{\boldsymbol\alpha}_k\,=\,+\,1\,. \label{ArightH}
\end{equation}
The fact that this ordered product yields a positive value confirms that
${\{{\boldsymbol\alpha}_x,\,{\boldsymbol\alpha}_y,\,{\boldsymbol\alpha}_z\}}$ indeed forms a right-handed frame of basis bivectors. This is a universally accepted convention, easily found in any textbook on geometric algebra.

Suppose now ${{\bf a}=a_{i}\,{\bf e}_i}$ and ${{\bf b}=b_{j}\,{\bf e}_j}$ are two unit vectors in ${{\rm I\!R}^3}$, expanded in right-handed basis ${\left\{{\bf e}_x,\,{\bf e}_y,\,{\bf e}_z\right\}}$, where the repeated indices are summed over ${x,\,y,}$ and ${z}$.
Then the right-handed set of graded basis defined in (\ref{Alisr}) leads to
\begin{equation}
\{\,a_i\;{\boldsymbol\alpha}_i\,\}\,\{\,b_j\;{\boldsymbol\alpha}_j\,\}\,=\,-\,a_i\,b_j\,\delta_{ij}\,-\,\epsilon_{ijk}\;a_i\,b_j\;{\boldsymbol\alpha}_k\,, \label{Abiiden-1}
\end{equation}
which, together with (\ref{Atog-13}) (which says that both ${+{\bf e}_i}$ and ${{\boldsymbol\alpha}_i}$ form right-handed frames), is equivalent to the identity
\begin{equation}
(\,I\cdot{\bf a})(\,I\cdot{\bf b})\,
=\,-\,{\bf a}\cdot{\bf b}\,-\,I\cdot({\bf a}\times{\bf b}), \label{Aone-ident}
\end{equation}
where ${I={{\bf e}_x}{{\bf e}_y}{{\bf e}_z}}$ is the standard trivector. Geometrically this identity describes all points of a parallelized 3-sphere.

Let us now consider a left-handed frame of ordered basis bivectors, which we denote by
${\{{\boldsymbol\beta}_x,\,{\boldsymbol\beta}_y,\,{\boldsymbol\beta}_z\}}$. It is important to recognize, however, that there is no {\it prior} way of knowing that this new basis frame is in fact left-handed. To ensure that it is indeed left-handed we must first make sure that it is an ordered frame by requiring that its basis elements satisfy the bivector properties analogous to those delineated in Eqs.${\,}$(\ref{Aprop13}) and (\ref{Aprop14}).
Next, to distinguish this frame\break from the right-handed frame defined by equation (\ref{ArightH}), we must require that its basis elements respect the property
\begin{equation}
{\boldsymbol\beta}_i\,{\boldsymbol\beta}_j\,{\boldsymbol\beta}_k\,=\,-\,1\,. \label{AleftH}
\end{equation}
One way to ensure this is by multiplying all vector and bivector elements in the basis set (\ref{Acl30}) by a minus sign, giving
\begin{equation}
\left\{1,\,\;-\,{\bf e}_x,\,-\,{\bf e}_y,\,-\,{\bf e}_z,\,\;
-\,{\bf e}_y\wedge{\bf e}_z,\,-\,{\bf e}_z\wedge{\bf e}_x,\,-\,{\bf e}_x\wedge{\bf e}_y,\,\;
{\bf e}_x\wedge{\bf e}_y\wedge{\bf e}_z\right\}\!. \label{Acl-right}
\end{equation}
As we shall soon see, this choice of the basis leads us to the incorrect result (\ref{Atwo-ident}) for our purposes, because we have not changed the sign of the pseudoscalar ${I:={\bf e}_x\wedge{\bf e}_y\wedge{\bf e}_z}$ in addition to that of the scalar 1. Consequently, although the four-dimensional even and odd subalgebras are now left-handed, the full eight-dimensional algebra remains right-handed, because we have changed the signs of only an {\it even} number of its elements, namely three vectors plus three bivectors, comprising six elements in total. Another way to see this is by noting that the determinant of the matrix that transforms the basis (\ref{Acl30}) into (\ref{Acl-right}) is ${(-1)^6=+1}$. On the other hand, instead of the relation (\ref{Atog-13}) we${\,}$now${\,}$have
\begin{equation}
{\boldsymbol\beta}_i\,=\,-\,{\bf e}_j\wedge{\bf e}_k\,=\,I\cdot(\,-\,{\bf e}_i\,)\,,\label{Atog-20}
\end{equation}
and therefore the condition (\ref{AleftH}) above is automatically satisfied. As is well known, this was the condition imposed by Hamilton on his unit quaternions, which we now know are nothing but a left-handed set of basis bivectors. It can be easily checked that the basis bivectors satisfying the properties (\ref{Aprop13}), (\ref{Aprop14}), (\ref{AleftH}), and (\ref{Atog-20}) compose the subalgebra
\begin{equation}
{\boldsymbol\beta}_i\,{\boldsymbol\beta}_j \,=\,-\,\delta_{ij}\,+\,\epsilon_{ijk}\,{\boldsymbol\beta}_k\,. \label{Alisr-left}
\end{equation}

Suppose now ${{\bf a}=a_{i}\,{\bf e}_i}$ and ${{\bf b}=b_{j}\,{\bf e}_j}$ are two unit vectors in ${{\rm I\!R}^3}$, identical to those used in Eq.${\,}$(\ref{Abiiden-1}), where the repeated indices are again summed over ${x,\,y,}$ and ${z}$. Then the left-handed set of graded basis defined in (\ref{Alisr-left})${\;}$leads${\;}$to
\begin{equation}
\{\,a_i\;{\boldsymbol\beta}_i\,\}\,\{\,b_j\;{\boldsymbol\beta}_j\,\}\,=\,-\,a_i\,b_j\,\delta_{ij}\,+\,\epsilon_{ijk}\;a_i\,b_j\;{\boldsymbol\beta}_k\,, \label{Abiiden-2}
\end{equation}
which, together with (\ref{Atog-20}) (which says that both ${-{\bf e}_i}$ and ${{\boldsymbol\beta}_i}$ form left-handed frames), is equivalent to the identity
\begin{equation}
(\,I\cdot{\bf a})(\,I\cdot{\bf b})\,
=\,-\,{\bf a}\cdot{\bf b}\,-\,I\cdot({\bf a}\times{\bf b}), \label{Atwo-ident}
\end{equation}
where ${I}$ is the standard trivector. But this is still a right-handed identity, giving a physically incorrect result for our purposes of experimentally distinguishing the left-handed bivectors from the right-handed bivectors unambiguously.

Indeed, the geometric identities (\ref{Aone-ident}) and (\ref{Atwo-ident}) are identical despite the fact that the corresponding bivectorial relations (\ref{Abiiden-1}) and (\ref{Abiiden-2}) are not. Thus, unlike the cross product, the geometric product between bivectors remains invariant under orientation changes if they are confined to the even ({\it i.e.}, bivector) and odd ({\it i.e.}, vector) subalgebras. But that is not at all surprising, because the relations (\ref{Atog-13}) and (\ref{Atog-20}) posit that the right-handed basis bivectors ${{\boldsymbol\alpha}_i}$ are dual to the right-handed basis vectors ${+{\bf e}_i}$ and the left-handed basis bivectors ${{\boldsymbol\beta}_i}$ are dual to the left-handed basis vectors ${-{\bf e}_i}$. However, in the EPR-Bohm type experiments one does not change coordinate systems back and forth to observe the spins. One keeps the vector basis ${\{\,+{\bf e}_i\,\}}$ fixed at both stations for the entire course of the experiment, in order to unambiguously determine whether a given spin is ``up'' or ``down'' about an experimentally fixed direction, represented by a specific vector. Therefore, to arrive at a genuinely left-handed counterpart of the identity (\ref{Aone-ident}) --- with\break all bivectors within both identities being dual to the vectors expanded in one and the same basis ${\{\,+{\bf e}_i\,\}}$, we must consider orientation changes in the entire algebra ${{Cl}_{3,0}}$ of the orthogonal directions in 3D space, by means of the basis
\begin{equation}
\left\{1,\,\;-\,{\bf e}_x,\,-\,{\bf e}_y,\,-\,{\bf e}_z,\,\;
-\,{\bf e}_y\wedge{\bf e}_z,\,-\,{\bf e}_z\wedge{\bf e}_x,\,-\,{\bf e}_x\wedge{\bf e}_y,\,\;
-\,{\bf e}_x\wedge{\bf e}_y\wedge{\bf e}_z\right\}\!, \label{Acl-left}
\end{equation}
where the sign of every non-scalar element is now different from that in the set (\ref{Acl30}), including the volume element ${I}$:
\begin{equation}
I \longrightarrow -I := -\,{\bf e}_x\wedge{\bf e}_y\wedge{\bf e}_z\,. \label{I}
\end{equation}
Since the determinant of the matrix that transforms (\ref{Acl30}) into (\ref{Acl-left}) is now ${(-1)^7=-1}$, the basis defined by (\ref{Acl-left}) are genuinely left-handed, {\it relative} to the basis defined by both (\ref{Acl30}) as well as (\ref{Acl-right}). The all important question then is: How do the identities (\ref{Aone-ident}) and (\ref{Atwo-ident}) transform under the handedness transformation (\ref{I}) of the volume element? But it is not difficult to see that under (\ref{I}) all three bivectors change signs and the identity (\ref{Atwo-ident}) transforms into
\begin{equation}
(\,I\cdot{\bf a})(\,I\cdot{\bf b})\,=\,-\,{\bf a}\cdot{\bf b}\,+\,I\cdot({\bf a}\times{\bf b}). \label{Athree-ident}
\end{equation}
Evidently, this is now a genuinely left-handed bivector identity compared to the right-handed identity (\ref{Aone-ident}). Crucially, there is a sign difference in the second term on the RHS of this identity, compared to the right-handed identity (\ref{Aone-ident}). Consequently, in perfect harmony with how the bivector relations (\ref{Abiiden-1}) and (\ref{Abiiden-2}) transform into one another under parity transformations, the identities (\ref{Aone-ident}) and (\ref{Athree-ident}) now transform into one another under the transformation (\ref{I}):
\begin{equation}
(\,I\cdot{\bf a})(\,I\cdot{\bf b})\,=\,-\,{\bf a}\cdot{\bf b}\,-\,I\cdot({\bf a}\times{\bf b})\,\xleftrightarrow{\;\;+I\,\longleftrightarrow\,-I\;\;\,}\,
(\,I\cdot{\bf a})(\,I\cdot{\bf b})\,=\,-\,{\bf a}\cdot{\bf b}\,+\,I\cdot({\bf a}\times{\bf b}).
\end{equation}
For convenience, we can now rewrite these two alternative identities (\ref{Aone-ident}) and (\ref{Athree-ident}) as two hidden variable possibilities
\begin{equation}
(\,+\,I\cdot{\bf a})(\,+\,I\cdot{\bf b})\,
=\,-\,{\bf a}\cdot{\bf b}\,-\,(\,+\,I\,)\cdot({\bf a}\times{\bf b}) \label{Aid-1}
\end{equation}
and
\begin{equation}
(\,-\,I\cdot{\bf a})(\,-\,I\cdot{\bf b})\,
=\,-\,{\bf a}\cdot{\bf b}\,-\,(\,-\,I\,)\cdot({\bf a}\times{\bf b}). \label{Aid-2}
\end{equation}
Exploiting the natural freedom of choice in characterizing the orientation of 3-sphere by either ${\,+\,I\,}$ or ${\,-\,I\,}$, we can now\break combine the identities (\ref{Aid-1}) and (\ref{Aid-2}) into a single hidden variable equation (at least for the computational purposes):
\begin{equation}
(\,\lambda\,I\cdot{\bf a})(\,\lambda\,I\cdot{\bf b})\,
=\,-\,{\bf a}\cdot{\bf b}\,-\,(\,\lambda\,I\,)\cdot({\bf a}\times{\bf b}), \label{Acombi-0}
\end{equation}
where ${\lambda=\pm\,1}$ now specifies the orientation of the 3-sphere.  It is important to keep in mind here that the combined equation (\ref{Acombi-0}) is simply a convenient shortcut for representing two completely independent initial states of the physical system, one corresponding to the counterclockwise orientation of the 3-sphere and the other corresponding to the clockwise orientation of the 3-sphere. Moreover, at no time are these two alternative possibilities mixed during the course of an experiment. They represent two independent physical scenarios, corresponding to two independent runs of the experiment. Next, if we employ the notation ${{\boldsymbol\mu}=\lambda\,I}$, then the combined identity (\ref{Acombi-0}) takes the convenient${\;}$form
\begin{equation}
(\,{\boldsymbol\mu}\cdot{\bf a})(\,{\boldsymbol\mu}\cdot{\bf b})\,
=\,-\,{\bf a}\cdot{\bf b}\,-\,{\boldsymbol\mu}\cdot({\bf a}\times{\bf b}).\label{Abi-identitzzzzzzzzzzzz}
\end{equation}
This is the central equation of my local model. I have used it in various forms and notations since 2007${\;}$\ocite{Adisproof}. It is simply\break an isomorphic representation of the familiar identity from quantum mechanics, with the correspondence ${\,{\boldsymbol\mu}\,\longleftrightarrow\, i{\boldsymbol\sigma}}$:
\begin{equation}
(i{\boldsymbol\sigma}\cdot{\bf a})(i{\boldsymbol\sigma}\cdot{\bf b})\,=\,-\,
{\bf a}\cdot{\bf b}\,\dbl\,-\,
i\,{\boldsymbol\sigma}\cdot({\bf a}\times{\bf b}).\label{Anowobserve}
\end{equation}

In Ref.${\,}$\ocite{local} I have combined the alternative bivector relations (\ref{Abiiden-1}) and (\ref{Abiiden-2}) into a single hidden${\;}$variable${\;}$equation
\begin{equation}
L_{\mu}(\lambda)\,L_{\nu}(\lambda) \,=\,-\,\delta_{\mu\nu}\,-\,\sum_{\rho}\,\epsilon_{\mu\nu\rho}\,L_{\rho}(\lambda)\,, \label{Awh-o8899}
\end{equation}
together with ${L_{\mu}(\lambda):=\lambda\,D_{\mu}}$, with alternative choices ${\lambda=\pm\,1}$ for the orientation of ${S^3}$. Contracting this equation on\break both sides with the components ${a^{\mu}}$ and ${b^{\nu}}$ of arbitrary unit vectors ${\bf a}$ and ${\bf b}$ then gives the combined bivector identity
\begin{equation}
{\bf L}({\bf a},\,\lambda)\,{\bf L}({\bf b},\,\lambda)\,=\,-\,{\bf a}\cdot{\bf b}\,-\,{\bf L}({\bf a}\times{\bf b},\,\lambda)\,, \label{A50}
\end{equation}
which is a convenient notation for the identity (\ref{Abi-identitzzzzzzzzzzzz}). It simply combines the two alternative identities (\ref{Aone-ident}) and (\ref{Athree-ident})\break into a single identity, rendering the unit bivector ${{\bf L}({\bf n},\,\lambda^k)}$ a random variable {\it relative} to the freely chosen fixed detector bivector ${{\bf D}({\bf n})}$, for a given run ${k}$:
\begin{equation}
{\bf L}({\bf n},\,\lambda^k)\,=\,\lambda^k\,{\bf D}({\bf n})\,\,\Longleftrightarrow\,\,{\bf D}({\bf n})\,=\,\lambda^k\,{\bf L}({\bf n},\,\lambda^k)\,. \label{A55}
\end{equation}
The expectation value of simultaneous outcomes ${{\mathscr A}({\bf a},\,{\lambda^k})=\pm1}$ and ${{\mathscr B}({\bf b},\,{\lambda^k})=\pm1}$ in ${S^3}$ then works out as follows: 
\begin{align}
{\cal E}({\bf a},\,{\bf b})\,&=\lim_{\,n\,\rightarrow\,\infty}\left[\frac{1}{n}\sum_{k\,=\,1}^{n}\,{\mathscr A}({\bf a},\,{\lambda}^k)\;{\mathscr B}({\bf b},\,{\lambda}^k)\right] \,=\lim_{\,n\,\rightarrow\,\infty}\left[\frac{1}{n}\sum_{k\,=\,1}^{n}\,{\bf L}({\bf a},\,\lambda^k)\,{\bf L}({\bf b},\,\lambda^k)\,\right] \nonumber \\
&=\frac{1}{2}\Big[{\bf L}({\bf a},\,{\lambda}^k=+1)\;{\bf L}({\bf b},\,{\lambda}^k=+1)\Big]\,+\,\frac{1}{2}\Big[{\bf L}({\bf a},\,{\lambda}^k=-1)\;{\bf L}({\bf b},\,{\lambda}^k=-1)\Big], \label{Aexp}
\end{align}
where the last simplification occurs because ${\lambda^k}$ is a fair coin. Using the relations (\ref{A50}) and (\ref{A55}) the above sum can now be evaluated by noting that the spins in the right and left oriented ${S^3}$ satisfy the following geometrical relations:
\begin{align}
{\bf L}({\bf a},\,{\lambda}^k=+1)\;{\bf L}({\bf b},\,{\lambda}^k=+1)\,&=\,-\,{\bf a}\cdot{\bf b}\,-\,{\bf L}({\bf a}\times{\bf b},\,{\lambda}^k=+1) \nonumber \\
&=\,-\,{\bf a}\cdot{\bf b}\,-\,{\bf D}({\bf a}\times{\bf b}) \nonumber \\
&=\,{\bf D}({\bf a})\;{\bf D}({\bf b})\,=\,(\,+\,I\cdot{\bf a})(\,+\,I\cdot{\bf b})
\end{align}
and
\begin{align}
{\bf L}({\bf a},\,{\lambda}^k=-1)\;{\bf L}({\bf b},\,{\lambda}^k=-1)\,&=\,-\,{\bf a}\cdot{\bf b}\,-\,{\bf L}({\bf a}\times{\bf b},\,{\lambda}^k=-1) \nonumber \\
&=\,-\,{\bf a}\cdot{\bf b}\,+\,{\bf D}({\bf a}\times{\bf b}) \nonumber \\
&=\,-\,{\bf b}\cdot{\bf a}\,-\,{\bf D}({\bf b}\times{\bf a}) \nonumber \\
&=\,{\bf D}({\bf b})\;{\bf D}({\bf a})\,=\,(\,+\,I\cdot{\bf b})(\,+\,I\cdot{\bf a}).
\end{align}
In other words, when ${\lambda^k}$ happens to be equal to ${+1}$, ${{\bf L}({\bf a},\,{\lambda}^k)\;{\bf L}({\bf b},\,{\lambda}^k)=(\,+\,I\cdot{\bf a})(\,+\,I\cdot{\bf b})}$, and when ${\lambda^k}$ happens to be equal to ${-1}$, ${\,{\bf L}({\bf a},\,{\lambda}^k)\;{\bf L}({\bf b},\,{\lambda}^k)=(\,+\,I\cdot{\bf b})(\,+\,I\cdot{\bf a})}$. Consequently, the expectation value (\ref{Aexp}) reduces at once to
\begin{equation}
{\cal E}({\bf a},\,{\bf b})\,=\,\frac{1}{2}(\,+\,I\cdot{\bf a})(\,+\,I\cdot{\bf b})\,+\,\frac{1}{2}(\,+\,I\cdot{\bf b})(\,+\,I\cdot{\bf a})\,
=\,-\,\frac{1}{2}\left\{{\bf a}{\bf b}\,+\,{\bf b}{\bf a}\right\}=\,-\,{\bf a}\cdot{\bf b}\,+\,0\,,\label{Astand-nossss}
\end{equation}
because the orientation ${\lambda^k}$ of ${S^3}$ is a fair coin. Here the last equality follows from the definition of the inner product.

\vfill\eject 

\renewcommand{\bibnumfmt}[1]{[{\bf B}#1]}
\renewcommand{\citenumfont}[1]{{\bf B}#1}

\section{\large Ordering Relation between the Spins and the Detectors as a Local Hidden Variable}\label{AppB12}

It is instructive to refute the arguments by Richard Gill somewhat differently. In the end the 3-sphere model for the EPR-Bohm correlation is a local hidden variable model. As such, we can simply define the hidden variable ${\lambda=\pm1}$ of the model as the ordering relation between the spin bivectors ${{\bf L}({\bf a},\,\lambda)}$ and ${{\bf L}({\bf b},\,\lambda)}$ and the detector bivectors ${{\bf D}({\bf a}})$ and ${{\bf D}({\bf b})}$, and only subsequently identify it with the orientation of the 3-sphere, as an equivalent version of this${\;}$definition.  

To this end, we again begin with the bivector subalgebra (\ref{Awh-o8899}) in the notation of Ref.${\,}$\ocite{local} as our central equation,
\begin{equation}
L_{\mu}(\lambda)\,L_{\nu}(\lambda) \,=\,-\,\delta_{\mu\nu}\,-\,\sum_{\rho}\,\epsilon_{\mu\nu\rho}\,L_{\rho}(\lambda)\,, \label{Awh-o889999}
\end{equation}
which, upon contraction of both sides with the components ${a^{\mu}}$ and ${b^{\nu}}$ of unit vectors ${\bf a}$ and ${\bf b}$ gives the bivector${\;}$identity
\begin{equation}
{\bf L}({\bf a},\,\lambda)\,{\bf L}({\bf b},\,\lambda)\,=\,-\,{\bf a}\cdot{\bf b}\,-\,{\bf L}({\bf a}\times{\bf b},\,\lambda)\,. \label{A5000}
\end{equation}
Next, instead of assuming the hidden variable ${\lambda=\pm1}$ to be an orientation of ${S^3}$, we define ${\lambda=\pm1}$ to be the ordering relation between the spin bivectors ${{\bf L}({\bf a},\,\lambda)}$ and ${{\bf L}({\bf b},\,\lambda)}$ and the detector bivectors ${{\bf D}({\bf a}})$ and ${{\bf D}({\bf b})}$, with ${50/50}$ chance:
\begin{equation}
{\bf L}({\bf a},\,{\lambda}=+1)\;{\bf L}({\bf b},\,{\lambda}=+1)\,=\,{\bf D}({\bf a})\;{\bf D}({\bf b}) \label{pair1}
\end{equation}
or
\begin{equation}
\text{\;}{\bf L}({\bf a},\,{\lambda}=-1)\;{\bf L}({\bf b},\,{\lambda}=-1)\,=\,{\bf D}({\bf b})\;{\bf D}({\bf a}). \label{pair2}
\end{equation}
Since the spins emerging from the source are oblivious to the detectors located at remote stations, we represent the spins with a trivector ${{\boldsymbol\mu}}$ and detectros with a trivector ${I}$, respectively, without assuming any relation between them:
\begin{equation}
{\bf L}({\bf n},\,{\lambda})\,=\,{\boldsymbol\mu}\cdot{\bf n} \label{JB}
\end{equation}
and
\begin{equation}
\text{\;}\;{\bf D}({\bf n})\,=\,I\cdot{\bf n}\,, \label{IB}
\end{equation}
for any given dual vector ${\bf n}$. Our intent now is to find the relationship between ${{\boldsymbol\mu}}$ and ${I}$ using the identities (\ref{A5000}) and 
\begin{equation}
{\bf D}({\bf a})\,{\bf D}({\bf b})\,=\,-\,{\bf a}\cdot{\bf b}\,-\,{\bf D}({\bf a}\times{\bf b})\,. \label{bbA5000}
\end{equation}
Substituting the right-hand sides of these identities into the ordering relations (\ref{pair1}) and (\ref{pair2}) reduces the relations to
\begin{equation}
-\,{\bf a}\cdot{\bf b}\,-\,{\bf L}({\bf a}\times{\bf b},\,\lambda=+1)\,=\,-\,{\bf a}\cdot{\bf b}\,-\,{\bf D}({\bf a}\times{\bf b})
\end{equation}
or
\begin{equation}
-\,{\bf a}\cdot{\bf b}\,-\,{\bf L}({\bf a}\times{\bf b},\,\lambda=-1)\,=\,-\,{\bf b}\cdot{\bf a}\,-\,{\bf D}({\bf b}\times{\bf a})\,=\,-\,{\bf a}\cdot{\bf b}\,+\,{\bf D}({\bf a}\times{\bf b})\,,
\end{equation}
which, after canceling the scalar factor ${-\,{\bf a}\cdot{\bf b}}$ and using ${\lambda=\pm1}$ and the definitions (\ref{JB}) and (\ref{IB}), further reduces${\;}$to
\begin{align}
{\bf L}({\bf a}\times{\bf b},\,\lambda)\,
&=\,\lambda\,{\bf D}({\bf a}\times{\bf b}) \\
{\boldsymbol\mu}\cdot({\bf a}\times{\bf b})\,
&=\,\lambda\,I\cdot({\bf a}\times{\bf b}) \\
{\boldsymbol\mu}\,
&=\,\lambda\,I\,. \label{B12}
\end{align}
We have thus proved that the ordering relations (\ref{pair1}) and (\ref{pair2}) between the spin bivectors ${{\bf L}({\bf a},\,\lambda)}$ and ${{\bf L}({\bf b},\,\lambda)}$ and the\break detector bivectors ${{\bf D}({\bf a}})$ and ${{\bf D}({\bf b})}$ are equivalent to our hypothesis that the orientation of the 3-sphere is a fair coin.

The expectation value of simultaneous outcomes ${{\mathscr A}({\bf a},\,{\lambda^k})=\pm1}$ and ${{\mathscr B}({\bf b},\,{\lambda^k})=\pm1}$ in ${S^3}$ then works out as before: 
\begin{align}
{\cal E}({\bf a},\,{\bf b})\,&=\lim_{\,n\,\rightarrow\,\infty}\left[\frac{1}{n}\sum_{k\,=\,1}^{n}\,{\mathscr A}({\bf a},\,{\lambda}^k)\;{\mathscr B}({\bf b},\,{\lambda}^k)\right] \,=\lim_{\,n\,\rightarrow\,\infty}\left[\frac{1}{n}\sum_{k\,=\,1}^{n}\,{\bf L}({\bf a},\,\lambda^k)\,{\bf L}({\bf b},\,\lambda^k)\,\right] \nonumber \\
&=\frac{1}{2}\Big[{\bf L}({\bf a},\,{\lambda}^k=+1)\;{\bf L}({\bf b},\,{\lambda}^k=+1)\Big]\,+\,\frac{1}{2}\Big[{\bf L}({\bf a},\,{\lambda}^k=-1)\;{\bf L}({\bf b},\,{\lambda}^k=-1)\Big] \nonumber \\
&=\,\frac{1}{2}\left\{D({\bf a})\,D({\bf b})\right\}+\frac{1}{2}\left\{D({\bf b})\,D({\bf a})\right\}\,=\,\frac{1}{2}(\,+\,I\cdot{\bf a})(\,+\,I\cdot{\bf b})\,+\,\frac{1}{2}(\,+\,I\cdot{\bf b})(\,+\,I\cdot{\bf a})\,=\,-\,{\bf a}\cdot{\bf b}\,,\label{nossssmouy}
\end{align}
because the ordering alternatives (\ref{pair1}) and (\ref{pair2}) are assumed to be a fair coin, and ${\,\frac{1}{2}\left\{{\bf a}{\bf b}+{\bf b}{\bf a}\right\}={\bf a}\cdot{\bf b}\,}$ by definition.

\vfill\eject 

\renewcommand{\bibnumfmt}[1]{[{\bf C}#1]}
\renewcommand{\citenumfont}[1]{{\bf C}#1}

\parskip 6pt 

\baselineskip 12.5pt

\section{\large Point-by-Point Refutal of Gill's Latest Critique of my Disproof of Bell's Theorem}

In the version 8 of his critique of my disproof, \ocite{Gill-1}, Richard D. Gill has changed the title of his earlier preprint and repeated his old arguments in somewhat different language, apart from adding some new {\it argumentum ad hominem} against me and against the editorial boards of the journals {\it Royal Society Open Science} and {\it IEEE Access} in which two of my three papers are recently published \ocite{RSOS}\ocite{Christian2014}. In addition, Gill has published a sequel to \ocite{Gill-1} on the arXiv, \ocite{Gill-socks}, in response to my latest paper \ocite{Christian2019b}. But, unfortunately, both of his critiques, \ocite{Gill-1} and \ocite{Gill-socks}, contain numerous mathematical and conceptual mistakes, similar to those in his earlier critiques. The trouble with \ocite{Gill-1} begins already with its new title, which reads:\,``Does geometric algebra provide a loophole to Bell's theorem?" But my disproof of Bell's theorem has nothing whatsoever to do with any loopholes. Thus, his new title reveals a lack of understanding of\break what my argument against Bell's theorem is all about. The contents of his critique reveal even more serious mistakes, and they multiply in his latest preprint \ocite{Gill-socks}. In what follows, I refute Gill's arguments in \ocite{Gill-socks} point-by-point, by first reproducing his own words and then giving my responses, in order to expose the elementary nature of his mistakes.

\begin{center}
\bf{My Response to What Gill Calls ``The Heart of the Matter"}:
\end{center}

\bigskip
\hrule
\smallskip

\noindent \underbar{Gill writes}:

This paper is a sequel to my paper \ocite{Gill-1} which analyses a whole sequence of papers by Joy Christian, including the publications \ocite{IJTP}, \ocite{RSOS},  \ocite{Christian2014}. Christian's ``oeuvre'' was recently extended with a pedagogical paper \ocite{Christian2019b}, and it is useful to similarly extend my own work.

\bigskip
\hrule
\smallskip

\noindent \underbar{My response}:

There is no ``extension" of Gill's ``work" in his latest critique. As we will soon see, it contains only repetitions of his previous mistakes and adds some new mistakes. But it does give an opportunity to expose his mistakes more clearly.

\bigskip
\hrule
\smallskip

\noindent \underbar{Gill writes}:

Equations (24) and (25) of \ocite{Christian2019b} talk of two sets of bivectors $\mathbf L(\mathbf a, \lambda)$ with the scalar $\lambda = +1$ and $\lambda = -1$ but these equations do not fix the relation between the two sets of bivectors. However that becomes fixed by equations (26) and (27). We learn that $\mathbf L(\mathbf a, \lambda) = \lambda I \mathbf a$ where $I$ is the trivector or pseudo-scalar $\mathbf e_1 \wedge \mathbf e_2 \wedge \mathbf e_3$, the context is Geometric Algebra, the main reference being the standard work Doran and Lasenby (2003) \ocite{DoranLasenby}. 
The symbols $\mathbf a$ and $\mathbf b$ stand for ordinary real 3D vectors. $I$ commutes with everything and $I^2= -1$. It follows that 
$$\mathbf L(\mathbf a, \lambda) \mathbf L(\mathbf b, \lambda) ~=~ \lambda^2 I^2 \mathbf a \mathbf b ~=~ - \mathbf a \mathbf b$$ which does not depend on $\lambda$ at all. 
\bigskip
\hrule
\smallskip

\noindent \underbar{My response}:

Equations (24) and (25) of \ocite{Christian2019b} do not talk of two sets of bivectors. They talk of only {\it one} set of spin bivectors $\mathbf L(\mathbf n, \lambda)$, with the choice of handedness $\lambda = +1$ or $\lambda = -1$ of the 3-sphere, {\it with respect to the detector bivectors} $\mathbf D(\mathbf n)$, where all\break vectors ${\bf n}$ are unit vectors. While the equation Gill has written is mathematically correct, its RHS is not independent of this {\it relative} handedness $\lambda$. It would be a meaningless equation if its LHS ``depended" on $\lambda$ but its RHS did not.

\bigskip
\hrule
\smallskip

\noindent \underbar{Gill writes}:

In fact, from geometric algebra we know that
$$-\, \mathbf a \mathbf b ~=~ - \mathbf a \cdot \mathbf b - I (\mathbf a \times \mathbf b) ~=~ - \mathbf a \cdot \mathbf b - \mathbf L(\mathbf a \times \mathbf b, +1).$$
So equation (25) is correct when $\lambda = +1$ but seems to be wrong when $\lambda = -1$.

\bigskip
\hrule
\smallskip

\noindent \underbar{My response}:

What Gill has missed here, and has missed it for the past eight years, is the fact that the handedness $\lambda$ of the spin bivector $\mathbf L(\mathbf n, \lambda)$ is meaningful only with respect to the handedness of the detector bivector $\mathbf D(\mathbf n)$, and vice versa. In fact, what we know from Geometric Algebra is not what Gill claims, but the following: 
\begin{equation}
-\, \mathbf a \mathbf b ~=~ - \mathbf a \cdot \mathbf b - \mathbf a \wedge \mathbf b ~=~ - \mathbf a \cdot \mathbf b - \mathbf L(\mathbf a \times \mathbf b,\,\lambda),
\end{equation}
where $\wedge$ represents the anti-symmetric outer product in GA. 
So equation (25) is correct for both $\lambda = +1$ and $\lambda = -1$. Once again, the handedness $\lambda$ of $\mathbf L(\mathbf n, \lambda)$ is meaningful {\it only} with respect to the handedness of $\mathbf D(\mathbf n)$, and vice versa. Gill's mistake here is to assume $\,\mathbf a \wedge \mathbf b = I (\mathbf a \times \mathbf b)\,$ instead of $\,\mathbf a \wedge \mathbf b = J(\mathbf a \times \mathbf b)$, with $J=\lambda I$ being the volume form on the 3-sphere of handedness $\lambda$. For further discussion, see Eq.~(\ref{B12}) of Appendix \ref{AppB12} above and Eq.~(98) of Ref.~\ocite{Christian2014}. 

\bigskip
\hrule
\smallskip

\noindent \underbar{Gill writes}:

Ah, but we now see where the handedness interpretation of $\lambda$ could come in. Perhaps the author has both a right-handed and a left-handed cross product. Introduce two cross-products, $\times\!(\lambda)$ where $\lambda =  \pm 1$, by the rules 
$$\mathbf  a \times\!\!(+1)\, \mathbf b ~=~ \mathbf a \times \mathbf b, \qquad \mathbf a \times\!\!(-1)\, \mathbf b ~=~ \mathbf b \times \mathbf a.$$ 

\hrule
\smallskip

\noindent \underbar{My response}:

The mistake made here by Gill is quite elementary. In vector algebra, the cross product between the vectors ${\bf a}$ and ${\bf b}$ is the same in both right-handed and left-handed coordinate frames. In the right-handed frame the three vectors, ${\bf a}$, ${\bf b}$, and ${{\bf a}\times{\bf b}}$, respect the right-hand rule and in the left-handed frame they respect the left-hand rule. But the rule for\break calculating the cross product remains the same. Therefore, the sign change in ${{\bf a}\times{\bf b}}$ introduced by Gill is illegitimate. 

Since Gill has exhibited much difficulty for the past eight years with this elementary concept from vector algebra, let me elaborate on it here further. Given the basis vectors ${{\bf e}_1}$, ${{\bf e}_2}$, and ${{\bf e}_3}$, in vector algebra the components and direction of the cross product vector ${\bf c}={{\bf a}\times{\bf b}}$ are calculated using the rules of calculating a determinant as follows:
\begin{equation}
{\bf c}\,=
\begin{vmatrix}
\;{\bf e}_1 & {\bf e}_2 & {\bf e}_3 \;\\ 
\;a_1 & a_2 & a_3 \\
\;b_1 & b_2 & b_3
\end{vmatrix}
=\,{\bf e}_1(a_2b_3-a_3b_2)+{\bf e}_2(a_3b_1-a_1b_3)+{\bf e}_3(a_1b_2-a_2b_1),
\end{equation}
where ${{\bf a}=a_1{\bf e}_1+a_2{\bf e}_2+a_3{\bf e}_3}$ and
${{\bf b}=b_1{\bf e}_1+b_2{\bf e}_2+b_3{\bf e}_3}$. Note that handedness does not enter this prescription at all. It is designed in such a way that the three vectors, ${\bf a}$, ${\bf b}$, and ${{\bf a}\times{\bf b}}$, in the given order, have the same handedness as the coordinate frame in which they are evaluated, regardless of the handedness of that frame  \ocite{Arfken}. In the left-handed coordinate frame the three vectors satisfy the left-hand rule, but the above prescription gives the same answer for ${\bf c}$.

This is not some quirk of vector algebra. There are very good reasons why the definition of a cross product has been devised in this manner. Vectors are ``geometric objects" \ocite{MTW}. They are so called because they exist independently of coordinate systems or reference frames. In particular, our vectors ${\bf a}$ and ${\bf b}$ and their cross product vector ${{\bf c} = {\bf a}\times{\bf b}}$ are all geometric objects. They all exist independently of coordinate systems or reference frames. Consequently, they remain independent of the handedness of a coordinate system or reference frame chosen to express them in practice.

\bigskip
\hrule
\smallskip

\noindent \underbar{Gill writes}:

Now equation (25), corrected, makes sense and is consistent with what follows:
$$\mathbf L(\mathbf a, \lambda) \mathbf L(\mathbf b, \lambda) ~=~ - \mathbf a \cdot \mathbf b - \mathbf L(\mathbf a \times\!\!(\lambda)\, \mathbf b, \lambda).$$
\hrule
\smallskip

\noindent \underbar{My response}:

Gill has not ``corrected" my equation (25) as he claims, but replaced it with his own incorrect equation, thereby introducing an illegitimate sign change in the definition of the cross product, just as he has done for the past eight years. The above equation harbors the same mistake he had made eight years ago in the second unnumbered equation of his critique \ocite{Gill-1}. As I have explained above [cf. Eq.~(\ref{bad-gill})], Gill has inserted an additional ${\lambda}$ in my equation, {\it by hand},\break in the middle of that equation. But, as we just saw, it does not belong there. I have explained this in the paragraph that includes Eq.~(36) in my rebuttal in \ocite{Response-3}. But this mistake in \ocite{Gill-1} remains uncorrected even after eight years.

To appreciate how nonsensical his version of my equation is, let us evaluate the bivector 
$\mathbf L(\mathbf a \times\!\!(\lambda)\, \mathbf b, \lambda)$ appearing on the RHS of his equation above:
\begin{equation}
\mathbf L(\mathbf a \times\!\!(\lambda)\, \mathbf b, \lambda) = \lambda I (\mathbf a \times\!\!(\lambda)\, \mathbf b)=\lambda^2 I(\mathbf a \times \mathbf b) =
I(\mathbf a \times \mathbf b)=\mathbf D(\mathbf a \times \mathbf b).
\end{equation}
Thus, Gill's version of my equation is actually this:
\begin{equation}
\mathbf L(\mathbf a, \lambda) \mathbf L(\mathbf b, \lambda) \,=\, -\, \mathbf a \cdot \mathbf b - \mathbf D(\mathbf a \times \mathbf b). \label{c4}
\end{equation}
So, in Gill's version of my equation, for $\lambda=-1$ the bivectors $\mathbf L(\mathbf a, \lambda)$ and $\mathbf L(\mathbf b, \lambda)$ appearing on its LHS are left-handed whereas the bivector $\mathbf D(\mathbf a \times \mathbf b)$ appearing on its RHS is right-handed. Thus, Gill's equation does not define the bivector subalgebra consistently in {\it any} coordinate frame. Far from restoring consistency, Gill has introduced {\it inconsistency}.

To see that Eq.~(\ref{c4}) above is the same as the second equation in the earlier versions of Gill's critique \ocite{Gill-1}, we use Eq.~(32) of my paper \ocite{Christian2019b} (which Gill agrees with) to recognize that $\mathbf D(\mathbf a \times \mathbf b) =\lambda\,\mathbf L(\mathbf a \times \mathbf b,\,\lambda)$, and rewrite (\ref{c4}) as
\begin{equation}
\mathbf L(\mathbf a, \lambda) \mathbf L(\mathbf b, \lambda) \,=\, -\, \mathbf a \cdot \mathbf b - \lambda\,\mathbf L(\mathbf a \times \mathbf b,\lambda). \label{c5}
\end{equation}
Note that this equation is not the same as my Eq.~(25). There is an extra $\lambda$ on the RHS that does not belong there. 

\bigskip
\hrule
\smallskip

\noindent \underbar{Gill writes}:

Having restored consistency to the definitions we can now quickly check formulas (30) and (31), taking account of (32), which give us $$\mathbf L(\mathbf a, \lambda) = \lambda I \mathbf a, \quad\mathbf D(\mathbf a) = \lambda \lambda I \mathbf a = I \mathbf a.$$

\hrule

\noindent \underbar{My response}:

As noted, Gill has not ``restored consistency" but introduced {\it inconsistency}. While the above two equations are correct, they are just my equation (32). They are not related to anything Gill has claimed in his previous comments. 

\bigskip
\hrule
\smallskip

\noindent \underbar{Gill writes}:

From the central expressions in (30) and (31) (the ones with limits as $\mathbf s_1$ converges to $\mathbf a$ and $\mathbf s_2$ converges to $\mathbf b$) we find that $$\mathcal A(\mathbf a, \lambda)= -\mathbf D(\mathbf a) \mathbf L (\mathbf a, \lambda) = - \lambda I^2 \mathbf a^2 = \lambda,$$
$$\mathcal B(\mathbf b, \lambda)= +\mathbf D(\mathbf b) \mathbf L (\mathbf b, \lambda) =  \lambda I^2 \mathbf b^2 =-\lambda,$$
exactly as the right hand sides of (30) and (31) proclaim.

\bigskip
\hrule
\smallskip

\noindent \underbar{My response}:

These equations are correct only for $\mathbf s_1 \not= \mathbf s_2$. They ignore the conservation of zero spin angular momentum, which amounts to
setting $\mathbf s_1 = \mathbf s_2$. In other word, the above equations hold in general only if the conservation of zero spin angular momentum is violated, or, equivalently, the M\"obius-like twists in the Hopf bundle of $S^3$ are ignored. That is to\break say, the equations hold if the entire argument of my paper is missed and one stoops back to the flat geometry of ${{\rm I\!R}^3}$.

\bigskip
\hrule
\smallskip

\noindent \underbar{Gill writes}:

This is consistent with (45)--(50). The final evaluation of (50) in (55)--(62) must be incorrect, the correlation must come out as $-1$. 

\bigskip
\hrule
\smallskip

\noindent \underbar{My response}:

It is easy to verify that the final evaluation of (50) in (55)--(62) is correct, and the correlation comes out as ${-{\bf a}\cdot{\bf b}}$. 

\bigskip
\hrule
\smallskip

\noindent \underbar{Gill writes}:

The author does not take account of the fact that surely, his cross-product too should also consistently follow the left- or right-handedness of the coordinate frame chosen by $\lambda$. 

\bigskip
\hrule
\smallskip

\noindent \underbar{My response}:

I have exposed Gill's mistake in this claim in my response above. It stems from his failure to understand how cross products are calculated in vector algebra. They do not change sign between left- and right-handed coordinate frames.

\bigskip
\hrule
\smallskip

\noindent \underbar{Gill writes}:

The interested reader may search for the mistake themselves, it is hidden in the derivation (34)--(40).

\bigskip
\hrule
\smallskip

\noindent \underbar{My response}:

It appears that either Gill has not been able to carry out the derivation (34)--(40) himself (even though it is quite straightforward), or has not been able to find any mistake in it. In fact, no invisible mistake exists in the derivation.

\underbar{Note added}: In v3 of \ocite{Gill-socks} Gill has rescinded his claim that there is a mistake ``hidden in the derivation (34)--(40)."

\bigskip
\hrule
\smallskip

\noindent \underbar{Gill writes}:

In the computer code in Section IV, the error in the evaluation of the correlation is ``fixed'' by the line 
$$\texttt{if(lambda==1) \{q=A B;\} else \{q=B A;\} }$$

\smallskip
\hrule
\smallskip

\noindent \underbar{My response}:

There is no error in the evaluation of the correlation to be ``fixed." The computer code correctly reflects what has been evaluated analytically. The errors are made by Gill himself for the past eight years, despite my patient teachings.

\bigskip
\hrule
\smallskip

\noindent \underbar{Gill writes}:

At the request of Joy Christian, I mention that he states that he has refuted all criticism of his works in the papers 
\ocite{Response-3}, \ocite{Response-5}, \ocite{Response-4}.

\bigskip
\hrule
\smallskip

\noindent \underbar{My response}:

I have also addressed various criticisms, and answered many questions in detail, in the Appendix B of Ref.~\ocite{Christian2014}.

\bigskip
\hrule
\bigskip

This completes my response to \ocite{Gill-socks}. But let me also reproduce here the summary of my responses to Gill's earlier criticisms. In the unpublished versions of \ocite{Gill-1} with a different title, Gill had attempted to criticize an earlier version of my local-realistic model presented in \ocite{frw}. His critique, however, contains some very surprising mathematical and conceptual mistakes. For example, the abstract of the first version of his preprint refers to the quantity ${-{\bf a}\cdot{\bf b}-{\bf a}\times{\bf b}}$ as a ``bivector." And even after my detailed explanations of the difference between a cross product and a wedge product, and the difference between a bivector and a multivector within geometric algebra, all subsequent versions of his preprint continue the mistake of referring to the multivector ${-{\bf a}\cdot{\bf b}-{\bf a}\wedge{\bf b}}$ as a bivector, leading to more serious mistakes later on in his critique. I have systematically corrected these mistakes in my responses \ocite{Response-3} and \ocite{Response-5}.

One of the surprising oversights in Gill's critique is the distinction between the detector bivectors ${{\bf D}({\bf a})}$ and ${{\bf D}({\bf b})}$ and the spin bivectors ${-{\bf L}({\bf s},\,\lambda)}$ and ${+{\bf L}({\bf s},\,\lambda)}$ considered in \ocite{frw}, together with the reciprocal relation between them,
\begin{equation}
{\bf L}({\bf n},\,\lambda)\,=\,\lambda\,{\bf D}({\bf n})\,\,\Longleftrightarrow\,\,{\bf D}({\bf n})\,=\,\lambda\,{\bf L}({\bf n},\,\lambda)\,,
\end{equation}
with ${\lambda}$ being the uncontrollable hidden variable in the sense of Bell \ocite{Bell-1964-666}. In other words, the correct representation of EPR-Bohm experiment and the corresponding spin detection processes defined in Eqs.~(58) and (59) are entirely missing in Gill's portrayal of my model. Consequently, what is described in the preprint \ocite{Gill-1} is {\it not} my model at all.

Moreover, Eq.~(4) of Gill's critique makes another serious mistake regarding the physics underlying the EPR-Bohm experiments. It inserts the equation ${{\mathscr A}({\bf a},\,\lambda){\mathscr B}({\bf b},\,\lambda)=(-\lambda)(+\lambda)=-1}$ for all ${\bf a}$ and ${\bf b}$ even when ${{\bf b}\not={\bf a}}$ by identifying ${{\mathscr A}({\bf a},\,\lambda)}$ with ${-\lambda}$ and ${{\mathscr B}({\bf b},\,\lambda)}$ with ${+\lambda}$, despite the fact that no such equation exists in my model. The insertion of this equation not only violates the conservation of spin angular momentum captured in Eqs.~(69) and (70) of \ocite{frw}, but also confuses the measurement results ${{\mathscr A}=\pm1}$ and ${{\mathscr B}=\pm1}$, which occur at remote stations, with the initial state ${\lambda=\pm1}$, which originate at the central source in the overlap of the backward light cones of Alice and Bob. It is\break evident from Eqs.~(69) and (70) that ${{\mathscr A}{\mathscr B}=-1}$ for ${{\bf b}\not={\bf a}}$ can occur if and only if the said conservation law is violated.

In summary, Gill's critique in all versions of \ocite{Gill-1} is a straw-man argument that ignores the fact that my approach to strong correlations is based on a {\it relative} orientation of a quaternionic 3-sphere, taken as Bell's local hidden variable. So much so, that Gill actually replaces one of my central equations with one of his own (thereby introducing a sign error), criticizes his own mistaken equation, and then declares that he has refuted my model. Indeed, in Eq.~(2) of his critique an additional ${\lambda}$ is inserted {\it by hand}, in the middle of that equation. As discussed above, I have explained this in the paragraph that includes Eq.~(36) in my response \ocite{Response-3}. But this mistake in \ocite{Gill-1} still remains uncorrected. 

What is more, in the latest version of \ocite{Gill-1} (version 8), Gill has added further elementary mistakes. For example, in the beginning of an argument he writes: ``Take any unit bivector $v$. It satisfies $v^2 = 1$ hence $v^2 - 1 = (v - 1)(v + 1) = 0$." But any unit bivector squares to $-1$, not $+1$. Consequently, this mistake reduces Gill's entire argument to absurdity.

In another paper \ocite{Gill-2} Gill criticizes my proposed experiment to test the relevance of Bell's theorem in a macroscopic setting \ocite{IJTP}. Unfortunately, this critique too contains surprisingly elementary mathematical and conceptual mistakes. For example, in the very equation of mine that this critique claims to be criticizing (namely, the standard definition of the bivector subalgebra \ocite{DoranLasenby}), Gill forgets to sum over the bivector-index, arriving at a rather strange conclusion. What is more, the Bell-CHSH correlator is also calculated incorrectly in \ocite{Gill-2}, by summing over spin detections of physically incompatible experiments. What is astonishing is that nowhere in my paper \ocite{IJTP} such a correlator involving incompatible experiments is even considered. In his critique, the correlator is simply made up, attributed to me, and then criticized. I have explained these and further errors in Gill's critiques in my response \ocite{Response-5} and analysis \ocite{Bell-oversight}.

\end{document}